\documentclass[a4paper,11pt]{article}
\pdfoutput=1 % if your are submitting a pdflatex (i.e. if you have
\usepackage[utf8]{inputenc}
\usepackage{jcappub} % for details on the use of the package, please
\usepackage{graphicx,rotating,wrapfig,float}
\usepackage{tabularx}
\usepackage{multirow}

\title{Measurement of the attenuation length of argon scintillation light in the ArDM LAr TPC}

\author[a]{J.~Calvo}
\author[a]{C.~Cantini}
\author[a]{P.~Crivelli}
\author[b]{M.~Daniel}
\author[a]{S.~Di~Luise}
\author[a]{A.~Gendotti}
\author[a]{S.~Horikawa}
\author[a]{L.~Molina-Bueno}
\author[b]{B.~Montes}
\author[a]{W.~Mu}
\author[a]{S.~Murphy}
\author[a]{G.~Natterer}
\author[a]{K.~Nguyen}
\author[a]{L.~Periale}
\author[a]{Y.~Quan}
\author[a]{B.~Radics}
\author[a]{C.~Regenfus}
\author[b]{L.~Romero}
\author[a]{A.~Rubbia}
\author[b]{R.~Santorelli}
\author[a]{F.~Sergiampietri}
\author[a]{T.~Viant}
\author[a]{S.~Wu}

\affiliation[a]{ETH Zurich, Institute for Particle Physics, Zurich, Switzerland}
\affiliation[b]{CIEMAT, Div. de F{\'\i}sica de Particulas, Avda. Complutense, 22, E-28040, Madrid, Spain}

\emailAdd{andre.rubbia@cern.ch}

\abstract{We report on a measurement of the attenuation length for the scintillation light in the tonne size liquid argon target of the \ardm\ dark matter experiment. The data was recorded in the first underground operation of the experiment in single-phase operational mode. The results were achieved by comparing the light yield spectra from \ar\ and \kr\ to a description of the \ardm\ setup with a model of full light ray tracing. A relatively low value close to 0.5\,m was found for the attenuation length of the liquid argon bulk to its own scintillation light. We interpret this result as a presence of optically active impurities in the liquid argon which are not filtered by the installed purification systems. We also present analyses of the argon gas employed for the filling and discuss cross sections in the vacuum ultraviolet of various molecules in respect to purity requirements in the context of large liquid argon installations.}

\newcommand{\gam}{\ensuremath{\gamma}}
\newcommand{\ardm}{ArDM}

\newcommand{\VUVabsL}{\ensuremath{\lambda_{\mathrm{VUV}}}}
\newcommand{\Refl}{\ensuremath{\mathcal{R}}}
\newcommand{\Rayl}{\ensuremath{l}}
\newcommand{\ardmrI}{ArDM Run\,I}

\newcommand{\creq}{\ensuremath{C_{\mathrm req}}}
\newcommand{\tslow}{\ensuremath{\tau_{\mathrm{slow}}}}
\newcommand{\tfast}{\ensuremath{\tau_{\mathrm{fast}}}}
\newcommand{\kr}{\ensuremath{\mathrm{^{83m}Kr}}}
\newcommand{\ar}{\ensuremath{\mathrm{^{39}Ar}}}

\newcommand{\HHO}{\ensuremath{\mathrm{H_{2}O}}}
\newcommand{\NN}{\ensuremath{\mathrm{N_{2}}}}
\newcommand{\OO}{\ensuremath{\mathrm{O_{2}}}}
\newcommand{\CHfour}{\ensuremath{\mathrm{CH_{4}}}}
\newcommand{\sigeff}{\ensuremath{\left<\sigma_{\mathrm eff}\right>}}
\newcommand{\meter}{\ensuremath{\,\mathrm{m}}}

\newcommand{\nm}{\ensuremath{\,\mathrm{nm}}}

\newcommand{\mus}{\ensuremath{\,\mu\mathrm{s}}}
\newcommand{\ns}{\ensuremath{\,\mathrm{ns}}}

\newcommand{\bm}{\boldmath}

\begin{document}

%%\linenumbers

\maketitle

\section{Introduction}

The scintillation light of liquid argon (LAr), emitted in the vacuum-ultraviolet (VUV) in a narrow band around 127\,nm, is well described by the formation and decay of argon excimers\,\cite{10.2307/77705,Suzuki79,:/content/aip/journal/jcp/91/3/10.1063/1.457108,0295-5075-91-6-62002}. In this work we investigate the influence of impurities on the VUV light yield in a tonne scale LAr target. In particular the following two processes are considered, the non-radiative destruction of excimers states, often referred-to as (impurity) quenching, and secondly and more important the absorption of produced VUV scintillation light during its propagation through the LAr bulk. By first principles both effects trace back to the presence of impurities in the argon. 

Collisional quenching of excimers by impurities is a well known phenomenon in LAr, becoming especially obvious due to the long lifetime of the triplet excimer state. The effect is determined by the two-body interaction rate which is competing with the radiative decay\,\cite{:/content/aip/journal/jcp/57/8/10.1063/1.1678779,1748-0221-5-05-P05003}, and leads to an apparent reduction of the emission time of the slow light component. As a consequence the light pulse shape, respectively the component ratio $CR$\,\footnote{for a definition see below}, change accordingly. It is possible to reconstruct the undisturbed shape from the literature value of the triplet lifetime in a pure argon environment\,\cite{1748-0221-3-02-P02001} and to determine the strength of this effect. Its exact quantitative estimation often lacks the knowledge of the collisional cross section. The effect of impurity quenching in LAr due to the varying concentrations of \OO , \NN\ and \CHfour\ can be found in literature\,\cite{1748-0221-5-05-P05003,1748-0221-5-06-P06003,Acciarri2009169, 1748-0221-8-12-P12015}. 

The absorption of LAr scintillation light propagating through the bulk is strictly bound to the presence of impurity states, since pure argon is fully transparent to its own scintillation light. Elastic interactions of VUV photons with argon atoms, e.g.~Rayleigh scattering, are possible, but only affect the direction of light propagation. The process of absorption is quantified by a wavelength dependent cross section of molecules, which is often referred to a molecular oscillator strength in spectroscopy. Due to the high density of the liquid argon, impurities at low concentrations (ppb) can have notable effects on the scintillation yield\,\cite{Neumeier201570}, above all in large detectors. Only a few elements dissolved in LAr (e.g.~\NN\ and \CHfour ) were systematically studied with respect to the attenuation length for LAr scintillation light\,\cite{1748-0221-8-07-P07011,1748-0221-8-12-P12015}, due to experimental difficulties (exponential law). No information on VUV absorption can be derived from the pulse shape study, since fast and slow components are equally affected. In measurements of light absorption a possible re-emission from the impurities has to be considered.

The VUV cross sections of a few elements diluted in LAr were compared to their pro\-per\-ties in the gaseous phase and found to be consistent with the room temperature results\,\cite{1748-0221-8-07-P07011,1748-0221-8-12-P12015}. A small shift and a broadening of the absorption lines was perceived in spectroscopic studies on elements embedded in liquid or solid argon. This can be attributed to the effect of the electronic band structure of the LAr\,\cite{RAZ1970511} and to its high density. By comparison with the gaseous phase, the main features of the spectra were found to be reproduced surprisingly well. One of the most intensively studied objects in dense argon are Xe atoms\,\cite{10.2307/77705,RAZ1970511,Neumeier:2012cz}. In Section\,\ref{sec:discussion} we discuss the extrapolation of these studies to  the estimation of molecular absorption cross sections of several molecules embedded in LAr and compare to our work. First of all, in Section\,\ref{sec:detectorsim}, we start to give an overview of the simulation framework of the \ardm\ experiment which is based on a description of the optical phenomena in the experiment by first principles. Section\,\ref{sec:data} presents the preparation of the data, which are analysed and compared to simulations in Section\,\ref{sec:characterisation}. In Section\,\ref{sec:bulkargon} and \ref{sec:qms} we discuss impurity trace analyses on the argon gas employed for filling the \ardm\ detector for its first underground operation. 

%%%%%%%%%%%%%%%%%%%%%%%%%%%%%%%%%%%%%%%%%%%%%%%%%%%%%%%%%%%%%%%%%%%%%%%%%%%%%%%%%%%%%%%%%%%%%%

\section{Simulation of the ArDM detector}
\label{sec:detectorsim}

The \ardm\ WIMP detector\cite{Rubbia:2005ge} consists of a 850\,kg active liquid argon target constructed as a vertical cylindrical TPC of 80\,cm diameter and 110\,cm in height. Scintillation light produced in the LAr is wavelength shifted on the side reflectors and diffused to two arrays of each 12 PMTs on top and bottom, which provide a photon counting light detection system (more details can be found in\,\cite{Rubbia:2005ge,Calvo:2015uln,thedetectorpaper}). In 2015 \ardm\ was operated for the first time at the Spanish underground lab Laboratorio Subterr\'aneo de Canfranc (LSC) with a full LAr target in the single-phase operational mode\,\cite{thedetectorpaper}.

\subsection{Simulation of the physics processes}
The response of the ArDM detector to particle interactions is calculated using the Geant4 toolkit\,\cite{Geant4}. All relevant particles (electrons, photons, neutrons and ions), described by a physics list, are considered for scattering off argon atoms or nuclei, taking also into account the subsequent physics phenomena. A custom made scintillation process was introduced for the simulation of the light emission for electron and nuclear recoil events in liquid and gaseous argon.

The scintillation light in gaseous and liquid argon is a result of the energy deposition by ionising particles leading to the production of excitons and free charges. Both contribute to the formation of excimers, in the first case via the well understood process of self-trapping of excitons, in the second via the recombination of electrons. The number of produced singlet and triplet excimer states and hence the intensities of the fast ($I_{\rm fast}$) and slow ($I_{\rm slow}$) components of the emitted VUV light depend on type and energy of the incident particle (ionisation density dependence)\,\footnote{For more details on the scintillation mechanism see e.g.~\,\cite{Suzuki79,Hitachi:1983zz,PhysRevLett.33.1365}}. %Sanami02 1748-0221-3-02-P02001  Suzuki82
The exploitation of this effect for the discrimination of background is particularly promising in LAr due to large difference of the lifetimes of the two excimer states. The light pulse shape can be parameterised by a component ratio $CR$ defined as $CR = I_{\rm fast}/(I_{\rm fast} + I_{\rm slow})$. Above roughly 50\,keV, the observed values for $CR$ are about 0.25 for minimal ionising particles (electron recoils), as well as about 0.75 for nuclear recoils and alpha particles\,\cite{Lippincott:2008ad,1742-6596-375-1-012019}. 

The \ardm\ simulation code uses the energy expenditure $W$\,=\,19.5\,eV\,\cite{Doke:2002} to calculate the average total number $\left<N\right>$ of VUV scintillation photons (of energy $E$\,=\,9.68\,eV) in liquid argon for a given energy deposit $E_{dep}$\,, i.e.~$\left<N\right> = E_{dep}/W$. The photon numbers of the fast and slow scintillation components are generated according to Poissonian PDFs and a value for $CR$, corresponding to the interaction type. The spatial position for the emission of VUV photons is uniformly distributed along the step length of the propagating particle. The time structure of the VUV emission is generated from a superposition of two exponential PDFs with the time constants of the fast and slow scintillation components; for \tfast\ a value of 7\,ns is employed \,\cite{Hitachi:1983zz}, while for \tslow\ the experimentally observed value of 1.23\mus\ is used\,\footnote{A value for \tslow, widely used in literature, is 1.6\mus\,\cite{Hitachi:1983zz}}.
The attenuation of the propagating VUV light is described by a simple exponential law with the characteristic parameter \VUVabsL , the VUV attenuation length. The wavelength shifting (WLS) efficiency of the thin TPB (tetraphenyl butadiene) coatings applied to the inner detector surfaces, is considered in the simulation by values ranging from 0.7 to 0.95, according to the different layer thicknesses, e.g.~for the side reflectors and PMT coatings. The emitted (optical) spectrum is centred around 435\,nm with a FWHM of about 180\,nm. Optical photons are isotropically emitted by the WLS and are propagated through the detector by ray tracing, applying all relevant physics processes (e.g.~reflection, scattering, refraction, total reflection). 

Both, the optical reflectivity \Refl\ of the main reflector foil, together with the attenuation length \VUVabsL , are the main parameters to be varied and estimated from data by the Bayesian fitting technique described in Section\,\ref{sec:characterisation}. To cross check results the Rayleigh scattering length \Rayl\ for VUV photons was varied in addition. In most part of this work a value of \Rayl\,=\,55\,cm (for LAr scintillation light) was chosen, taken from literature\,\cite{Grace15}.

\subsection{Simulation of the detector components}

\begin{figure}[hbtp]
\begin{center}
\includegraphics[width=0.8\textwidth]{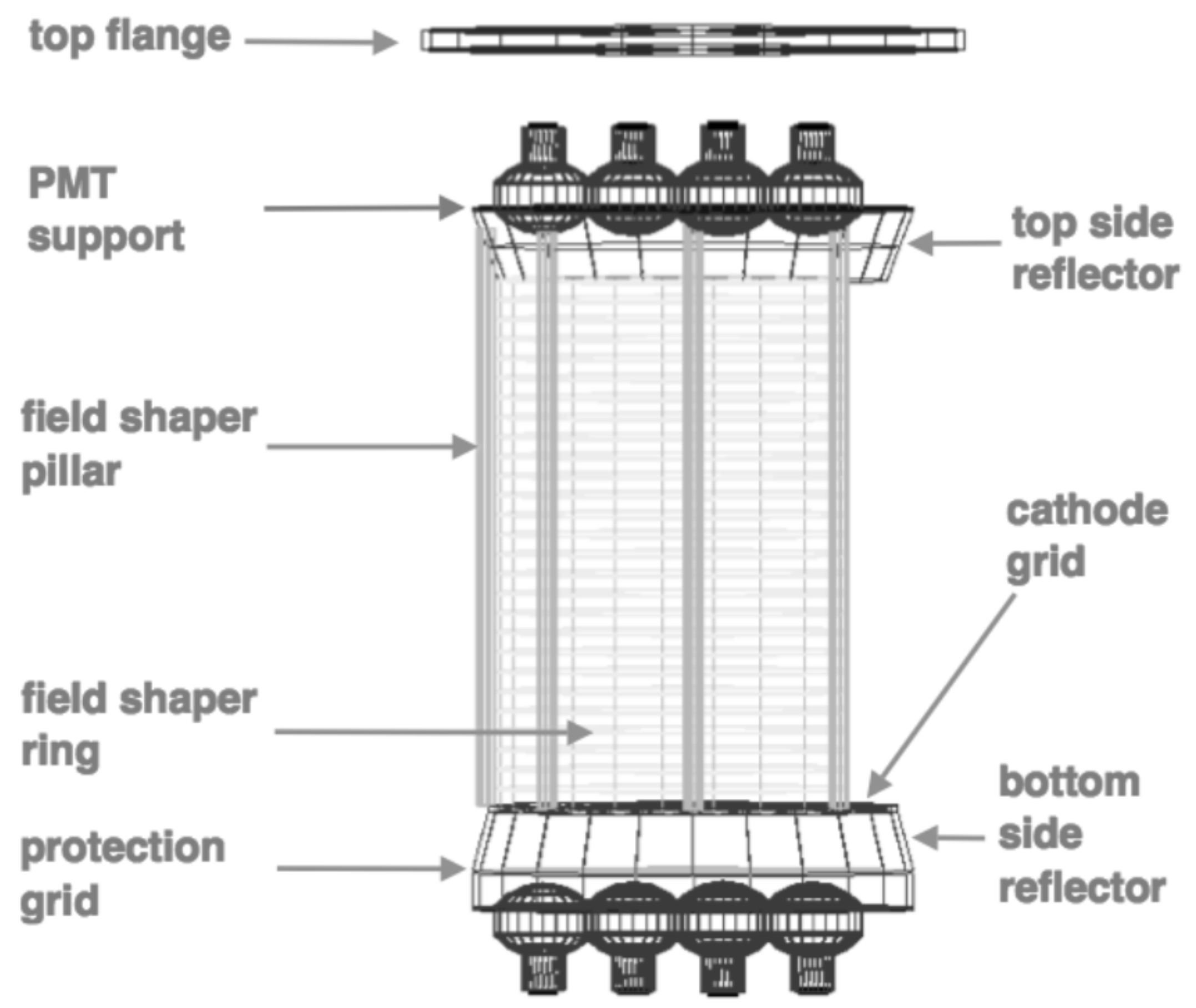}
\caption{The simulated geometry of the ArDM experiment.}
\label{fig:DetSim}
\end{center}
\end{figure}

In the simulation large emphasis was given to a detailed description of the material budgets of the individual detector components to correctly describe particle interactions with the experiment, as well as the effect of radiation emitted from radioactive traces in materials on the neutron and $\gamma$-backgrounds. The geometry of the ArDM experiment, as it is described in the simulation, is illustrated in Fig~\ref{fig:DetSim}, showing the inner detector components. The surrounding main experimental dewar (not shown) containing the LAr target has a multi-layer structure of stainless steel walls, LAr cooling layers, as well as vacuum sections. For simplification it is described by a single stainless steel layer of the same amount of material (of about 20\,mm thickness).

The 4 cm thick stainless steel top flange of the experiment contains a number of pillars reserved for cabling and cryogenic purposes. In the simulation the pillars are not included, but the effective amount of steel is accounted for by adjusting the thickness of the top flange. 

Two arrays of photomultiplier tubes (PMT), placed on the top and bottom of the main dewar, provide the light detection for the experiment. The spherical geometry of the PMT body is described using simple geometrical objects available in Geant4. The PMT surface contains, going from outside to inside, the WLS layer made of TPB, a PMT glass window and a thin photo cathode. After emission of a visible photon in the WLS material the photon may undergo various optical boundary processes along the LAr - WLS, WLS - PMT glass, PMT glass - PMT cathode boundaries. In the simulation a $d_{TPB} \simeq 10 \mu$m is used for the thickness of the TPB layer. The optical surfaces on the LAr - WLS, WLS - PMT boundaries are defined as non-absorbing dielectric-dielectric surfaces with 100\% Lambertian reflection mode. Both VUV and visible photons can pass through the LAr - WLS surface, but VUV light is fully absorbed on the WLS - PMT glass surface. No optical surface is defined for the PMT glass - cathode boundary. All photons are absorbed on this interface. In the simulation the PMT cathodes are the sensitive detectors of the experiment. Optical photons are detected with a detection probability depending on the hit position and impinging angle, given by the manufacturer (Hamamatsu).

The field shaper rings, mounted to vertical pillars, are arranged to generate a homogeneous electric drift field in the ArDM detector. Presently the field shaping system consists of 27 hollow stainless steel rings supported by 7 high density polyethylene pillars. In the work described here no electric fields were applied for data taking and simulations.

The main and side reflectors, constructed from a sandwich of Tetratex (Donaldson membranes) and a multilayer reflector foil (Vikuity ESR, 3M), are mounted onto the inner sides of the field shaper rings and are also coated with TPB. The optical boundary between the TPB coating and the Tetratex foil is defined to be dielectric and 100\% Lambertian for visible light. The value for its reflectivity \Refl\ is roughly known from measurements in the lab and subject to studies described in this work.  

Cathode and protection grids, each about 95\% transparent, are located at the bottom of the main reflector, with the protection grid placed 13\,cm below the cathode grid. The interface to the LAr is defined to be dielectric - metal and 100\% Lambertian. The reflectivity is set to a value of 50\% , estimated from measurements in the lab.

\subsection{Simulation of the PMT single photoelectron response}

The response of the PMTs to a single photo-electron ($pe$) is approximated from the measured PMT parameters by two Gaussian PDFs with $\sigma_{\rm t}$\,=\,3\ns\ for the temporal response, as well as $\sigma_{\rm ph}\,\simeq\,0.35\,\cdot{\left<pe\right>}$ for pulse height fluctuations; 
%\begin{equation}
%V(t) = V_{0}\exp\left(-\frac{(t-t_0)^{2}}{2\sigma_{\rm t}^{2}}\right)~and~V_0(ph) = \left<pe\right>\exp\left(-\frac{(ph-)^{2}}{2\sigma_{\rm t}^{2}}\right)
%\end{equation}
$t_0$ is related to the arrival time of a detected photon on the PMT cathode and ${\left<pe\right>}$ corresponds to the mean single photo-electron charge. The exact values for $\sigma_{\rm ph}$ are taken individually for each of the 24 PMTs from a look-up table obtained by calibration. This was done by adjusting the gains of all PMTs varying the HV settings to have the same value for ${\left<pe\right>}$ in all channels. The Gaussian shapes approximate the photo-electron pulse responses closely. 

Noise is generated in the ArDM simulation according to the measured dark count rates and electronic pedestal fluctuations (white noise). PMT dark counts occur randomly and uniformly during data taking due to thermal emission of electrons from either the cathode or one of the dynodes. The average dark count rate in the acquisition window is 0.01\,pe/4$\mu s$. Electronic white noise is simulated based on the Gaussian spread of pedestal distribution, obtained from measurement. The signal over noise ratio for single photo-electrons is of the order of $S/N \simeq 40$.

%%%%%%%%%%%%%%%%%%%%%%%%%%%%%%%%%%%%%%%%%%%%%%%%%%%%%%%%%%%%%%%%%%%%%%%%%%%%%%%%%%%%%%%%%%%%%%
\section{Preparation of data sets}
\label{sec:data}

A total of 3.3 billion triggers were recorded during ArDM Run I. About 10\% of the data was collected during calibration runs with radioactive test sources. All data was triggered under the same conditions, a signal above $\sim$2\,$pe$, in either the top, or bottom PMT array.
% ($pe$ = mean single photo electron signal).
With the full size LAr target the trigger rate amounted to roughly 1.3\,kHz, mainly due to $\beta$-decays of \ar\ in the LAr target. About 10\% of the triggers exhibit a much faster time structure than signals created by argon scintillation light. Such events are characterised by small signal amplitudes typical for single photon events fluctuating above the 2\,$pe$ trigger level. A fraction of those events contain higher amplitudes (up to $\sim$15\,$pe$) implying other origins than thermal emission of electrons in the photo cathodes, e.g.~after-pulsing or \v{C}erenkov light from the glass of the PMT windows. These events can easily be discriminated and do not present serious background to the data. 

The analyses presented in this paper compare the measured energy spectra from two data sets with Monte Carlo simulations. The first set was prepared from data collected without any calibration sources to study the \ar\ $\beta$-spectrum (endpoint 565\,keV). The second set consists of calibration data from \kr\ atoms (from a $^{83}$Rb source) injected into the gas phase of the detector. The metastable atoms (mean life 1.8\,h) deposit most of their de-excitation energy (41.5\,keV) via emissions of Conversion- and Auger-electrons. % \cite{Manalaysay:2009yq}. 
Technical details about the \kr\ source installed in the ArDM setup can be found in\,\cite{thedetectorpaper}. The \ar\ and \kr\ data sets were prepared by selecting events of the electron recoil type (ERL) and a rough fiducialisation cut, as explained in the following. 

Light yields are calculated by finding and summing clusters of photon signals over the acquisition time of 4\mus . The light yields $L_{\rm top}$ and $L_{\rm btm}$ are calculated from the sum of all signal clusters found in the top and bottom PMTs, respectively. The total light yield is calculated from the sum of both, $L_{\rm tot} = L_{\rm top} + L_{\rm btm}$. A vertical localisation parameter $TTR$ (top-to-total ratio) is introduced from the ratio of the top to the total yield, $TTR = L_{\rm top} /L_{\rm tot}$. This value is related to the vertical position of the event and used for fiducialisation cuts. We also define a pulse shape parameter $f90$, the ratio of the light detected in the first 90\,ns of the event to the total yield. A cut on $f90<1$ removes about 10\% of the events, which originate purely in noise. The final data selection is done by the conditions 0.2$\,<\,$f90$\,<\,$0.5 ($f90$-cut) and   0.2$\,<\,$TTR$\,<\,$0.8 ($TTR$-cut). In addition thresholds of 4\,$pe$ are applied for both, $L_{\rm top}$ and $L_{\rm btm}$, corresponding to a total energy threshold of about 8\,keV. The remaining fraction of events after applying these cuts amounts to 70\% of the events selected by the $f90<1$ condition, or hence 63\% of the triggered data.

The same cuts are applied to the simulated data sets of pure \ar\ and \kr\ events to evaluate the selection efficiencies. For each of them and each set of optical parameters, 200\,k events were generated uniformly distributed over the volume of the active LAr target. Here the $f90$-cut rejects $<$2\% of the \ar\ events and about 11\% of the \kr\ events due to larger statistical fluctuations at the lower energies. About 20\% of the events are removed by the $TTR$-cut in both data sets. The overall selection efficiencies in the MC data sets are $78.5\,\pm\,0.1$\% and $69.6\,\pm\,0.1$\% for the \ar\ and \kr , respectively. \gam\ background in the data is not derived from simulation but parameterised by an exponential distribution (see Section \ref{sec:bayesian-method}).

%%%%%%%%%%%%%%%%%%%%%%%%%%%%%%%%%%%%%%%%%%%%%%%%%%%%%%%%%%%%%%%%%%%%%%%%%%%%%%%%%%%%%%%%%%%%%%

\section{Measured light yield}
\label{sec:characterisation}

The light yield is determined by the response of the detector to the primary scintillation light. It can be parametrised by a set of optical parameters of the Monte-Carlo model. The exact values of the parameters are a priori not precisely known and need to be determined from the data. From the large set of parameters in the model, we found that a few most strongly affect the detector response. In particular, the attenuation length \VUVabsL\ of the argon scintillation light in the liquid argon itself, and the reflectivity \Refl\ of the reflector foil (in the following we refer to this parameter as ``reflectivity'') are most critical. In a preliminary analysis of Run\,I data a relatively small value far below 1\,m was found for \VUVabsL . Modelling the data by a short Rayleigh scattering length could not reproduce the data. On the other hand, a lower limit of 110\,cm for the attenuation length of very pure (distilled) liquid argon to its own scintillation light was recently found by experiment\,\cite{Neumeier201570}. The low value observed in our setup might be an indication for the presence of traces of impurities in the argon. The relatively short attenuation length in our target is also affecting the overall light yield, which was found smaller than expected\,\cite{thedetectorpaper}. For future runs of ArDM it is therefore critical to verify and understand the origin of the observed effect.

In order to obtain a quantitative estimation of the VUV light attenuation length \VUVabsL ,  the measured light yields of \ar\ and \kr\ events were compared to Monte Carlo data sets, in which \VUVabsL , as well as the reflectivity \Refl\ were scanned over the ranges of 40 - 200\,cm and 83 - 99\% respectively. The estimates for both parameters describing best our data were evaluated using a Bayesian likelihood maximisation technique described below. The Rayleigh scattering length was kept fixed at a value of 55\,cm.
In the simulation \ar\ beta electrons were generated in a range of $E = [3\,\mathrm{keV}, 600\,\mathrm{keV}]$ according to the theoretical spectrum described by the phase space factors, the Fermi correction, as well as the first forbidden Gamow-Teller transitions\,\cite{Bettini07,Konopinski66,Daniel68,Keefer04}. \kr\ events were generated by electrons of a fixed energy of $E = 41.5$ keV. Events were homogeneously distributed in the liquid argon volume. 
%\cite{Kastens:2009pa}. 

\subsection{Bayesian parameter estimation method}
\label{sec:bayesian-method}

The level of compatibility between data and simulation with a given set of parameters was evaluated with a Bayesian method, using the Bayesian Analysis Toolkit (BAT)\cite{Caldwell20092197}. The BAT framework allows to construct parametrised functional or template based models (likelihood models) from theory or simulation, performing the numerical evaluation of the Bayes-theorem on a data set $\vec{D}$ as follows,
\begin{equation}\label{eq:bayes}
p(\vec{a}|\vec{D}) = \frac{p(\vec{D}|\vec{a})p(\vec{a})}{p(\vec{D})} ,
\end{equation}
where $p(\vec{a}|\vec{D})$ denotes the conditional probability of the parameter set $\vec{a}$ given the data $\vec{D}$, $p(\vec{D}|\vec{a})$ denotes the conditional probability of the data set given a set of parameters $\vec{a}$, and finally $p(\vec{a})$ and $p(\vec{D})$ denote the probabilities of the parameter set and that of the data, respectively. 

The conditional probability from the right-hand side of the Bayes-equation was evaluated using the ArDM Monte Carlo simulation. We allowed for uniform prior probability for each parameter value in the (\VUVabsL, \Refl ) parameter space. In our Monte Carlo template based likelihood model we constructed the conditional probability explicitly as,
\begin{equation}\label{eq:likelihood}
p(\vec{D}|\vec{a}) = \prod\limits_{i} \frac{1}{\sqrt{2\pi}\sigma}e^{-\frac{(y_{i} - f(x_{i};\vec{a}))^{2}}{2\sigma^{2}}}
\end{equation}
where the product index $i$ runs through all the data bins of the observable used for the evaluation. In a given data bin, $i$, the likelihood model assumes bin entries with Gaussian uncertainties parametrised as the discrepancy between the data bin entries, $y_i$, and the Monte Carlo simulation entries in the same bin, $f(x_i; \vec{a})$. The Monte Carlo prediction for the number of events in bin $i$ depends on the particular parameter set (\VUVabsL, \Refl ), and a linear interpolation was used to obtain the approximate distributions between the parameter points. Light yield distributions in multiple TTR slices were used as the observable for the evaluation. Various TTR values correspond to different parts of the detector, and therefore BAT performs a combined evaluation using information from different parts of the ArDM detector together. We also allow for an overall Monte Carlo scaling factor, $A$, to normalize the events to the data. 

In the case of \ar\ data analysis the likelihood model was constructed using the Geant4 \ar\ simulation with an additional exponential term, with parameters $B$ and $C$, for the description of the gamma background in the high energy region. Therefore in the Bayesian evaluation of the \ar\ data the following function was used for the light signal prediction in bin $i$
\begin{equation}\label{eq:MCfuncAr39}
f(x_i;\vec{a}) = A\cdot g_{\mathrm{MC}}(x_i;\VUVabsL ,\Refl , \ar) + e^{(B + C\cdot x_{i})}.
\end{equation}
Here $g_{\mathrm{MC}}(x_i;\VUVabsL , \Refl, \ar )$ indicates the light signal prediction from the Monte Carlo simulation of \ar\ decays in bin $i$ for the parameter values \VUVabsL\ and \Refl . A lower threshold of 200\,$pe$ was used for the \ar\ evaluation to avoid a trigger bias or problems due to low energy backgrounds.
% Kr83 Likelihood model:

For the likelihood model of the \kr\  data the Geant4 \kr\ simulation was used together with the background from dedicated ArDM runs taken without Kr. The following model was used,
\begin{equation}\label{eq:MCfuncKr83}
f(x_i;\vec{a}) = A\cdot g_{\mathrm{MC}}(x_i;\VUVabsL, \Refl , \kr ) + h_{\mathrm{bkg}}(x_i),
\end{equation}
where $g_{\mathrm{MC}}(x_i;\VUVabsL ,\Refl, \kr )$ is the light signal prediction from Monte Carlo simulation of \kr , and $h_{\mathrm{bkg}}(x_i)$
is the contribution from \ardmrI\ background runs. The background was normalised to the data using the data acquisition time. In this case the signal dominates the low energy region with a maximum at around 40\,$pe$. A lower boundary of 20\,$pe$ was used in the evaluation, which was varied for the estimation of systematic uncertainties.

\begin{figure}[htb]
\centering
\includegraphics[height=0.6\textwidth]{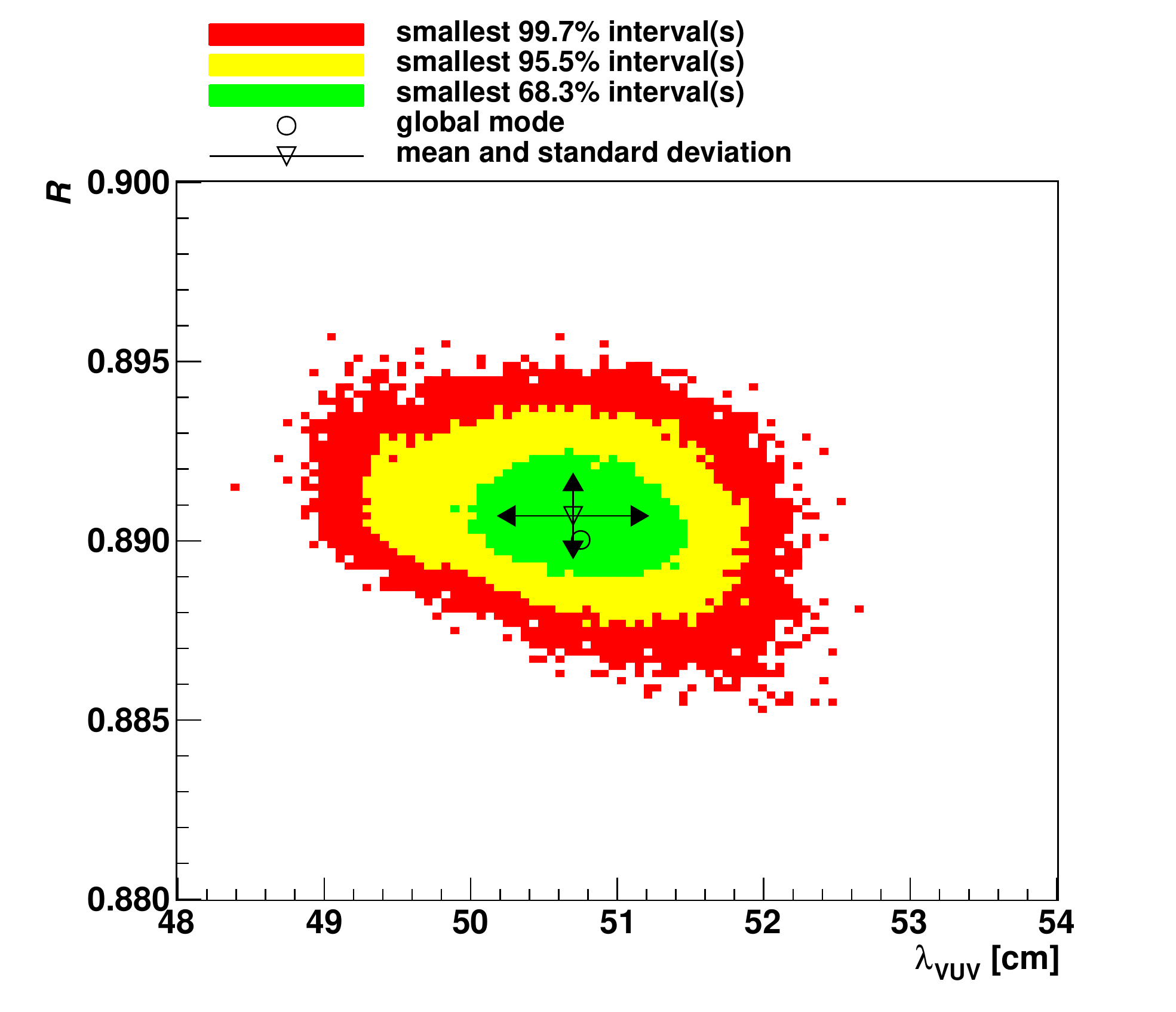}
\caption{Marginalized posterior distribution for the parameters \VUVabsL\ and \Refl\ from the Bayesian analysis output using $^{39}$Ar with exponential model to fit the data. The colour scale indicates the 68\% (green), 95.5\% (yellow) and 99.7\% (red) area around the maximum.   }
\label{fig:BAT_marg}
\end{figure}

The BAT evalution starts with a search for the global maximum of the full posterior. Two separate algorithms were tried (Minuit MIGRAD and Simulated Annealing, see \cite{Caldwell20092197} for further details) to check for the stability of the global maximum, both gave consistent results. During the analysis first the full parameter range of  \VUVabsL\,=\,[40\,cm, 200\,cm] and \Refl\,=\,[83\%, 99\%] was given to BAT. Then once the global maximum was established the BAT fit was repeated with the parameter range gradually reduced to be around the global maximum for large statistics sampling. As an illustration of the BAT output the marginalized posterior distribution for \VUVabsL\ and \Refl\ is shown from one of the $^{39}$Ar evaluations on Fig~\ref{fig:BAT_marg}. Similar results have been obtained for the \kr\ analysis. The maximum in the marginalized distribution matches closely to the global maximum of the full posterior. 

\subsection{Results}
\label{sec:results}

The fitting technique described above was applied to multiple groups of the data from \ardmrI . Results were found to be consistent over the entire run period spanning over more than 6 months. The estimation of the parameters at the global maximum for the likelihood, together with systematic uncertainties are shown in Table\,\ref{tab:BayesPars} for both, the \ar\ and \kr\ data sets. Results agree on values of around 50\,cm for the attenuation length \VUVabsL\ and about 90$\%$ for the reflectivity \Refl . Systematic errors were obtained by variation of the energy threshold, as well as the LY scale. The former, changing from 100 to 300\,$pe$ for the \ar , and 10 to 30\,$pe$ for the \kr\ data sets, produce uncertainties for \VUVabsL\ of 10 and 5\,cm, respectively. An overall 5\% error for the reflectivity is found. 
\begin{table}[ht]
\begin{center}
	\begin{tabular}{|l|c|c|c|c|} \hline
	   {\bf Results for:} & {\bf\bm \VUVabsL\ (\ar )} & 
	   {\bf\bm \VUVabsL\ (\kr )} & {\bf\bm \Refl\ (\ar )} & {\bf\bm \Refl\ (\kr )}  \\ \hline 
	  {\bf -\,global maximum} & 52.1 cm & 53.6 cm & 88.6\% & 92.4\% \\ \hline
	  {\bf -\,threshold variations}  & 49.4, 58.8 cm & 53.4, 57.3 cm& 86.8, 89.5 \%   &92.5, 91.2\%  \\ \hline
	   \bf\bm{-\,LY scale $\pm 10 \%$ } & 47.9, 57.7 cm & 54.3, 55.4 cm & 86.5, 90.2 \% &  88.0, 95.2\% \\ \hline
	\end{tabular}  
	\caption{Estimated VUV attenuation lengths as well as optical reflectivities from the output of the Bayesian analysis toolkit for the\ar\ and \kr\ data sets at maximum probability (Global maximum). The results for systematic variations of the threshold, as well as the light yield scale, are also shown (see text).}
\label{tab:BayesPars}
\end{center}
\end{table}
Varying the light yield scale by 10\% (up and down) creates a similar size of uncertainty for the attenuation length with almost no impact on \Refl\ in the case of \ar , while in the case of \kr\ this variation impacts mainly the reflectivity by about 7\%.  Therefore we presume  an overall systematic uncertainty of 20\% for the values of \VUVabsL\ and \Refl . The stability of fit results in respect to different calibration campaigns was cross-checked for data taken 24 hours before and after the \kr\ injections. As mentioned above a smeared \ar\ beta spectrum and an exponential parametrisation of the background was used for this purpose.   

\begin{figure}[htb]
\centering
\includegraphics[height=0.33\textwidth]{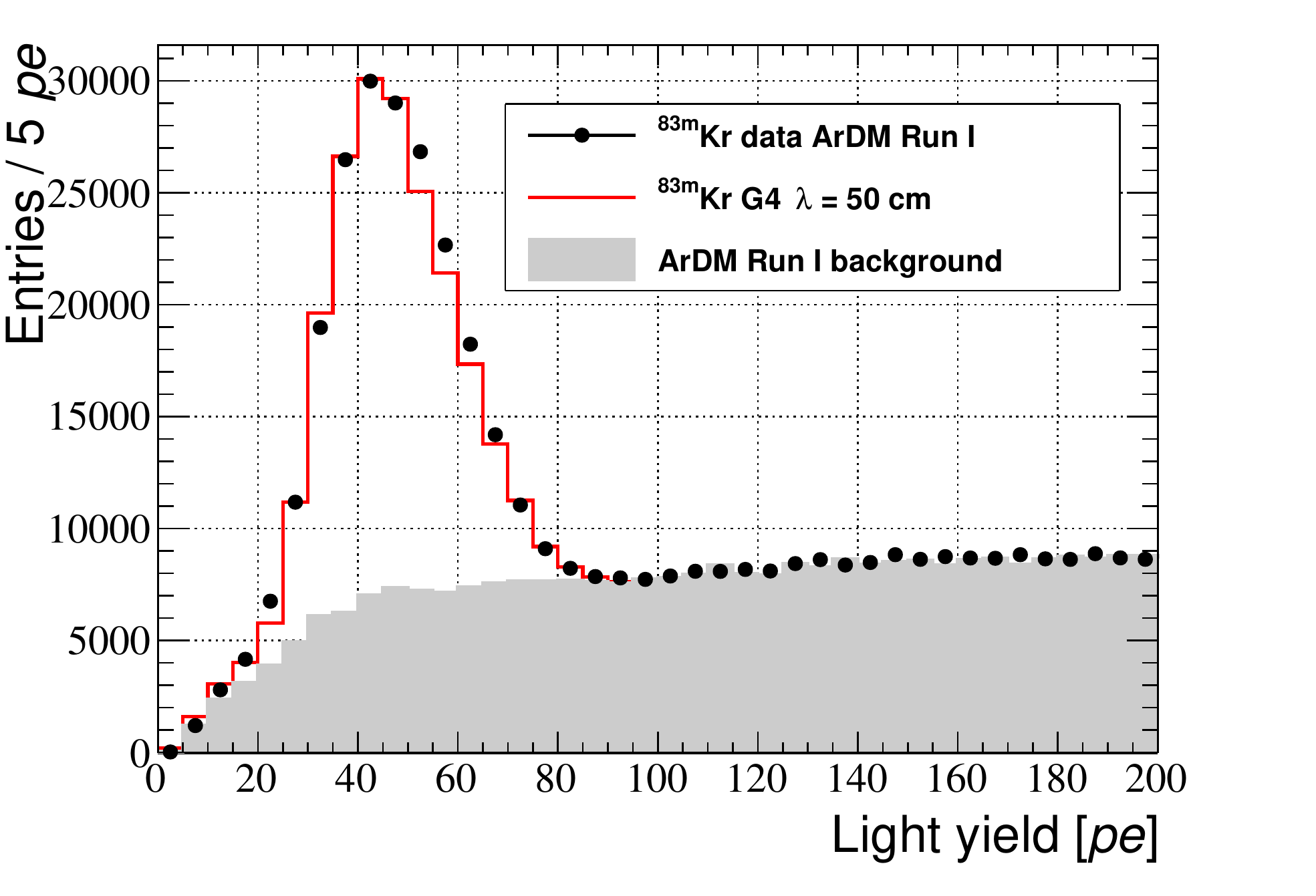}
\includegraphics[height=0.33\textwidth]{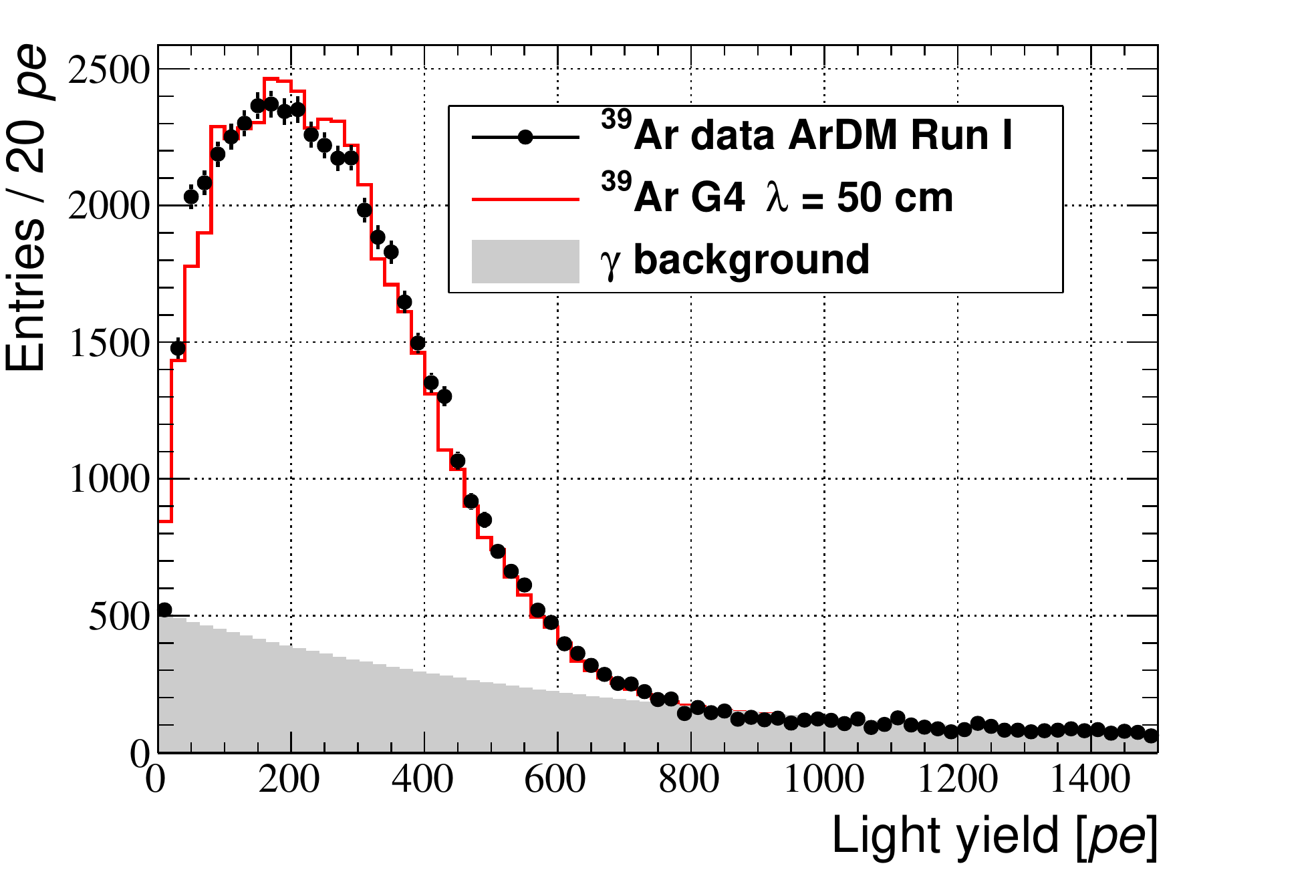}
\caption{Data (black dots) superimposed to Geant4 spectra of \kr\ (left) and \ar\ (right) events using the best tuned parameter set. Gray areas depict the main backgrounds for the two data sets.}
\label{fig:MCdata_bestfit}
\end{figure}

The generally good agreement between data and Geant4 simulations is illustrated in Fig~\ref{fig:MCdata_bestfit} for the \kr\ (left) and the \ar\ (right) data sets using parameters estimated at the global maximum. The light yield spectra are shown for a single TTR slice. Data is drawn by black dots while the Geant4 simulations are shown in red. 
The backgrounds which were used in the spectral fits are shown as gray shaded areas. They were derived from data taken before and after the injection of \kr\ atoms (left), as well as the tail of the \ar\ spectrum (exponential) above 800\,$pe$ (right). 

\begin{figure}[htb]
\centering
\includegraphics[height=0.33\textwidth]{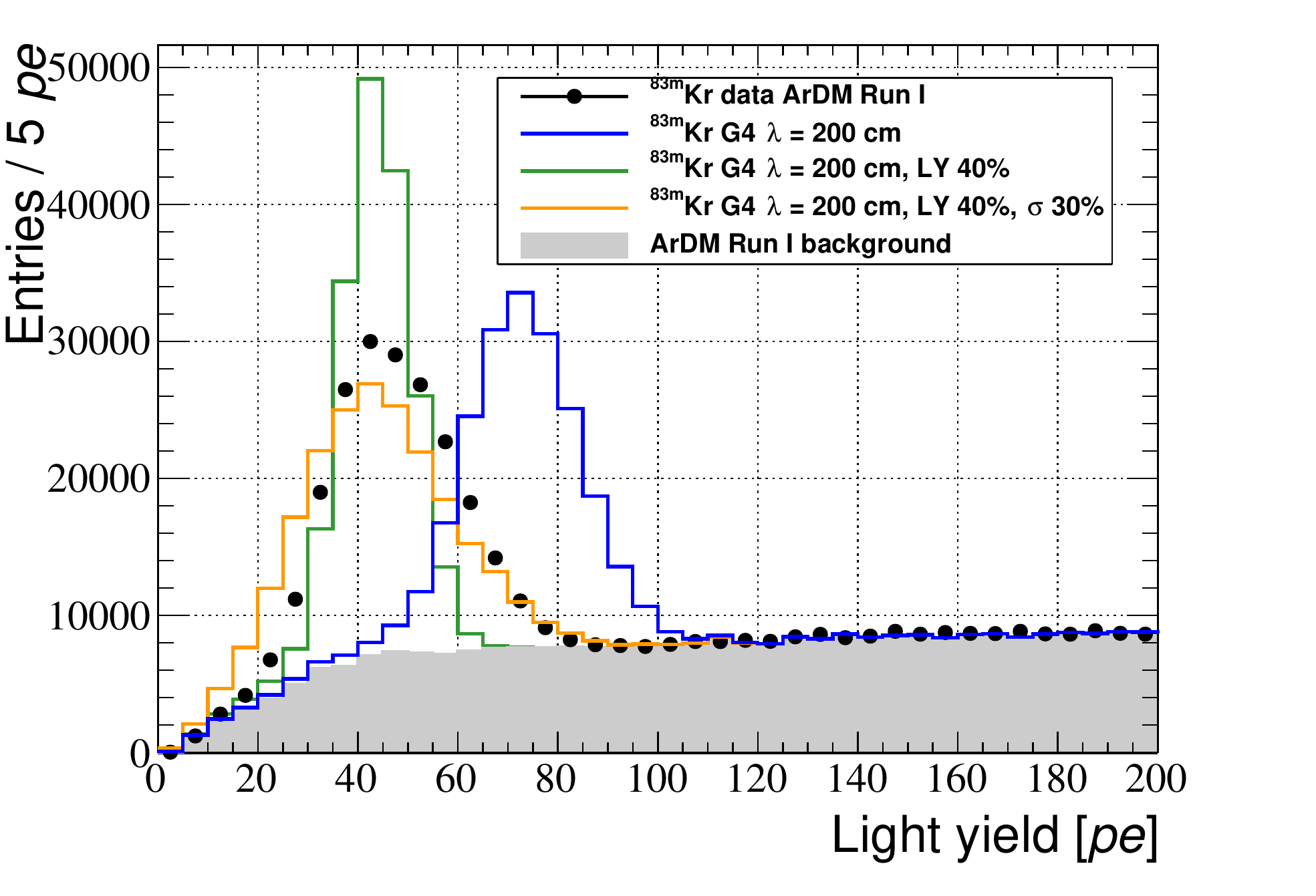}
\includegraphics[height=0.33\textwidth]{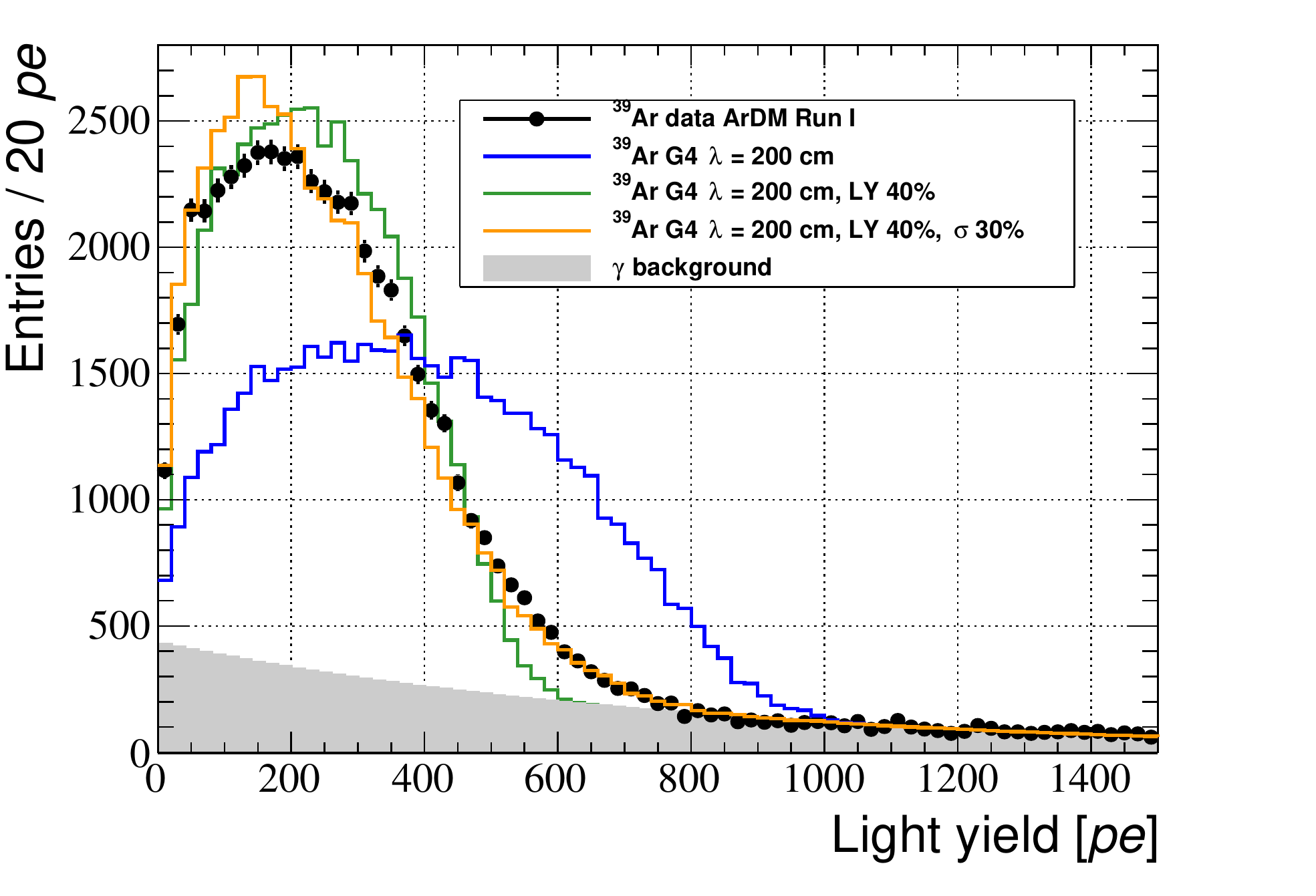}
\caption{Data (black dots) compared to MC spectra of \ar . The best tuned parameter set (\VUVabsL\,=\,55\,cm) are shown on the left. On the right plot  \VUVabsL\,=\,200\,cm with the set of best-tuned parameters
is shown in black. A spectrum with an ad hoc 40\% reduction in light yield scale (red) and with smearing by additional 30\% (brown) are also shown.}
\label{fig:mc_abs}
\end{figure}

To scrutinise the result of the short attenuation length we compare the data to simulated \ar\ and \kr\ spectra generated with a value of 200\,cm for \VUVabsL . Figure\,\ref{fig:mc_abs} shows again the data (black dots) with solid lines representing different stages of scaling and smearing of the simulated spectra. From data generated in this way (blue) we scale the light yield by an ad hoc value of 40\% (green). 
Obviously the rescaled spectra can reproduce the average yield, both for \kr\ and \ar\,, but distributions are much narrower than in the data. This becomes particularly evident in the region between 500 and 800~$pe$ for the \ar\ spectrum. An additional ad hoc energy smearing of 30\% (yellow) needs to be added to reproduce the widths of the distributions. 

This exercise indicates the presence of a spatially non-uniform process causing a reduction and smearing of the light signal in the experiment. A short attenuation length for VUV light traveling to a wavelength shifting surface, as well as efficient transportation of the shifted light to the PMTs, reproduce this effect perfectly. We want to stress that this observation became notable to such an extent only due to the large size of the liquid argon target of the ArDM experiment.

In summary, the observed light yields and shapes of the \ar\ and \kr\ spectra suggest a relatively short value of around 50\,cm for the VUV attenuation length in the LAr target for the first data taken underground at the \ardm\ experiment.  An hypothesis of the description of the data with an expected value (e.g.~200\,cm) for \VUVabsL\ and an unrealisticly low value for the reflectivity of the main reflector is clearly disfavoured. This result is also supported by the analysis of the gas data which can be found elsewhere\,\cite{thedetectorpaper}. A description of the data with a much reduced value for the Rayleigh scattering length, also does not fit the data. We interpret this result as the presence of optically active impurities in the liquid argon target which seem not to be filtered by the installed purification systems (see Ref.\cite{thedetectorpaper} for details on the purification process).

%%%%%%%%%%%%%%%%%%%%%%%%%%%%%%%%%%%%%%%%%%%%%%%%%%%%%%%%%%%%%%%%%%%%%%%%%%
\section{Discussion of the result in respect to molecular absorption cross sections}
\label{sec:discussion}

\begin{figure}[htb]
\centering
\includegraphics[height=0.60\textwidth]{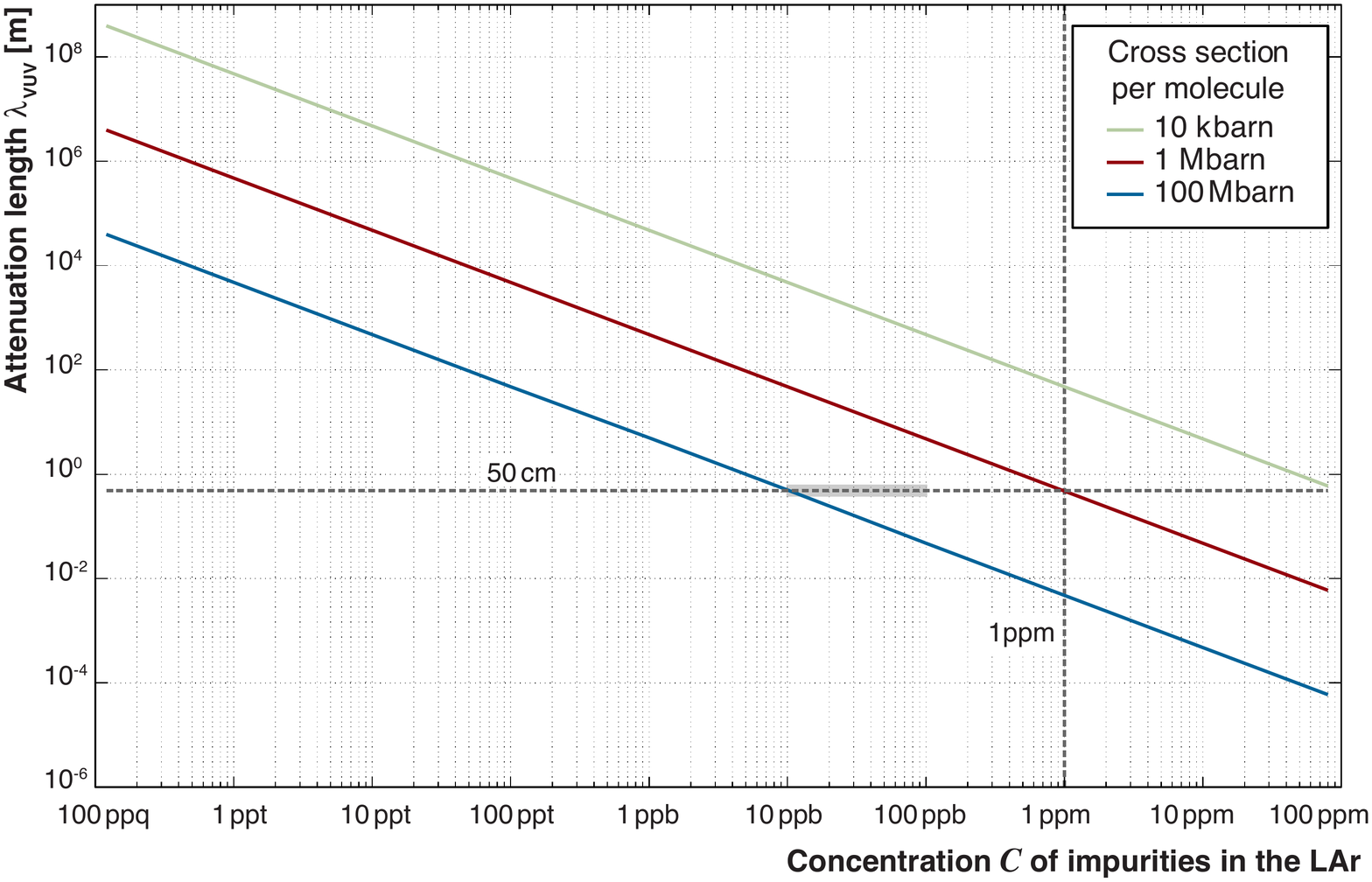}
%\caption{VUV absorption cross section per molecule vs.~their concentration in the LAr. The lines represent equal attenuation lengths.}
\caption{VUV attenuation length in dependence of the concentration of impurities in LAr. The diagonal lines represent molecular cross sections of 0.01, 1 and 100\,Mbarn respectively. The horizontal dotted line illustrates the value of \VUVabsL\,=\,0.5\,m.}
\label{fig:LabsVsConc}
\end{figure}

In this section we estimate the VUV attenuation length for impurities dissolved in LAr in dependence of a molecular absorption cross section $\sigma$ and compare to our result. Later on we evaluate cross sections of several relevant molecules in the spectral range of LAr scintillation light. Figure\,\ref{fig:LabsVsConc} shows the calculated attenuation length \VUVabsL, at a given concentration $C$. The 3 diagonal lines illustrate the cases for $\sigma$\,=\,0.01, 1 and 100\,Mbarn. The attenuation length was estimated from the total density of impurities,
\begin{equation}
\VUVabsL\,\,=\,\,\frac{1}{\sigma\cdot C\cdot n}\,\,\simeq\,\,\frac{476\meter}{\sigma\,[{\rm Mbarn}]\cdot C\,[{\rm ppb}]}
\label{equ:absL}
\end{equation}
where $n=2.1\cdot\,10^{28}\,m^{-3}$ is the density of atoms of liquid argon. The horizontal dotted line illustrates the value of 0.5\,m for the attenuation length found in this work. The vertical dotted line (1ppm) gives the upper limit for the sum of all impurities contained in the argon gas employed in the experiment, 
as quoted by the manufacturer\,\footnote{ALPHAGAZ-2 Argon, Air Liquide, \url{https://www.airliquide.com/}} for argon N60, 99.9999\% pure. In particular an upper limit of 100\,ppb is specified for the individual components of \OO , C$_{\rm n}$H$_{\rm m}$, CO, CO$_2$ and H$_2$, as well as 500\,ppb for \HHO\ and 300\,ppb for \NN , respectively. Water (and oxygen) traces are filtered-out by the installed purification system, but are not explicitly monitored. Due to its low cross section, nitrogen can be ruled out as affecting strongly the VUV absorption in LAr (see Table\,\ref{tab:cs}). The grey bar at \VUVabsL\,=\,50\,cm in Fig.\,\ref{fig:LabsVsConc} indicates the most probable range for impurity concentrations for this work, assuming maximal levels of 100\,ppb. Hence a photoabsorption cross section above  10\,Mbarn is required to affect significantly the VUV attenuation length in the LAr target of \ardm .

As mentioned above the Xe atom is a well studied object with respect to its oscillator strength in a dense argon environment\,\cite{10.2307/77705,RAZ1970511,Neumeier:2012cz}. This is especially interesting for the fact of the vicinity of its spectral lines to the emission spectrum of LAr. Very comparable results were found for the absorption spectra of free and in argon embedded Xe atoms. E.g.~the well known atomic transitions $^1S_0\rightarrow\,^3P_1$ (146.9\,nm) and $^1S_0\rightarrow\,^1P_1$ (129.7\,nm) appear as broadened and blue shifted (by 10 and 7\,nm) lines for Xe atoms diluted in solid and liquid argon host matrices respectively\,\cite{10.2307/77705}. In the same work a similarly strong, additional absorption line at 127\,nm was discovered, which was attributed to a Wanier-Mott type exciton state formed by the interaction of the Xe with the band structure of the liquid argon. This observation was confirmed in spectroscopic studies of VUV attenuation of LAr\,\cite{Neumeier:2012cz} and exposes Xe as a primary candidate for
\begin{figure}[htb]
\centering
\includegraphics[height=0.70\textwidth]{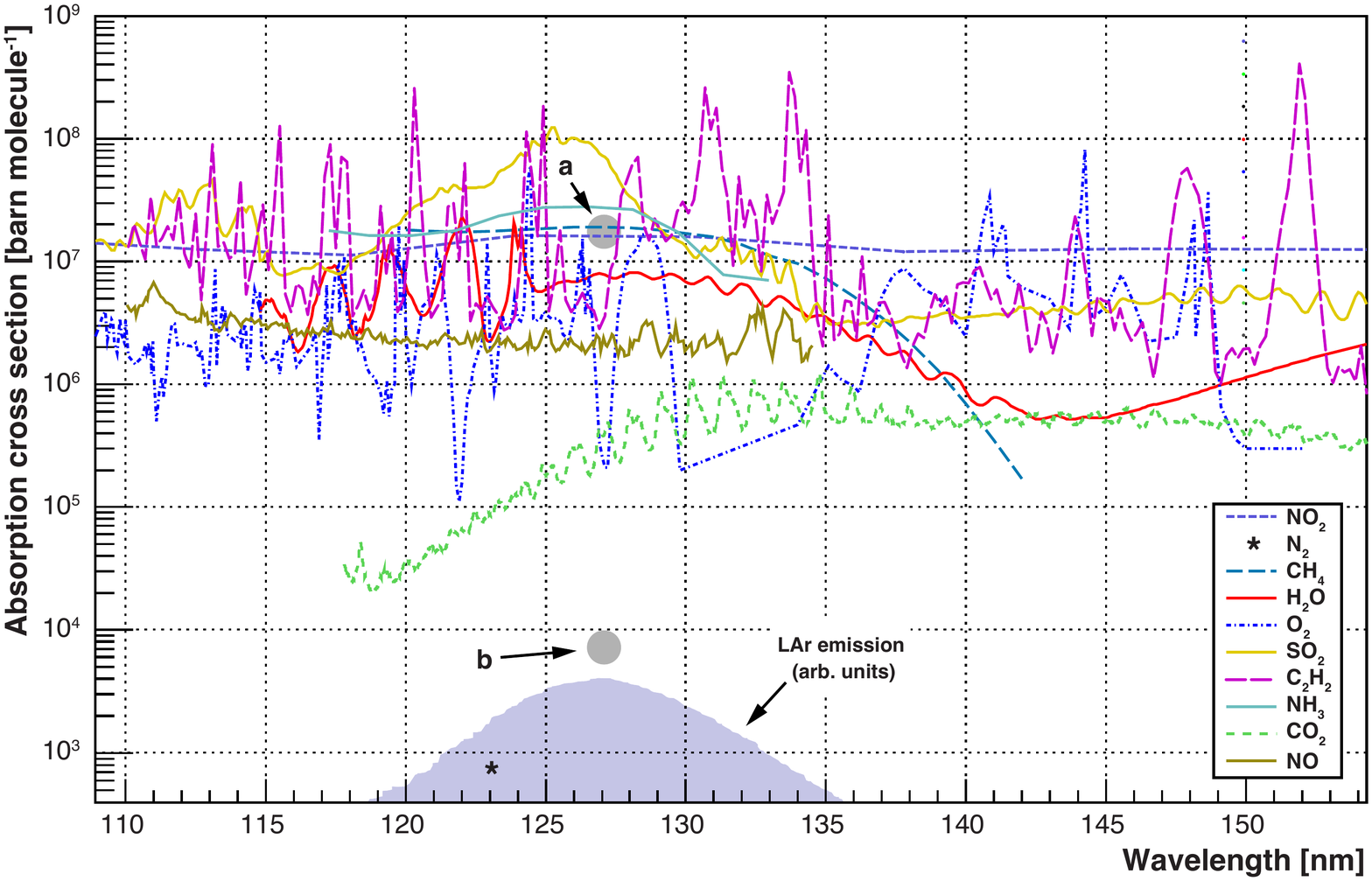}
\caption{Absorption cross sections of different molecules in gaseous phase as a function of the incident wavelength. The grey circles a and b represent the estimated cross sections of \CHfour\,\cite{1748-0221-8-12-P12015} and \NN\,\cite{1748-0221-8-07-P07011} diluted in LAr. The LAr emission spectrum is taken from\,\cite{Neumeier:2012cz} and plotted in grey (arb.~units).}
\label{fig:AbsXsections}
\end{figure}
absorbing LAr scintillation light. From the same work an effective cross section of 35\,Mbarn was estimated from the width of the absorption spectra around 127\,nm\,\cite{AndreasUlrich}, consistent with the value of about 25\,Mbarn derived from spectroscopic measurements on Xe gas\,\cite{PhysRevA.46.149}. 

Further evidence for the valid extrapolation of photoabsoprtion properties from the gaseous phase to the one in LAr, is derived from two direct measurements of VUV absorption of LAr bulk doped with different quantities of the absorber under investigation. The (approximate) cross sections of 15\,Mbarn and 7\,kbarn were determined for (a) \CHfour\,\cite{1748-0221-8-12-P12015} and (b) \NN\,\cite{1748-0221-8-07-P07011}, consistent with measurements done in gaseous \CHfour\ and \NN . Figure\,\ref{fig:AbsXsections} shows these values as grey dots together with several selected molecular VUV absorption cross sections relevant for this work (plotted as solid and dashed lines). The data was taken from the web site of MPI/Mainz\,\cite{spectral_atlas}. For \NN\ only one data point (at 123\,nm) was available and is marked by an asterisk. The low values (and scarce data) are due to a large drop in the oscillator strength between the ionisation edge at 14.5\,eV (85\,nm) and onset of the atomic lines at around 400\,\nm. Figure\,\ref{fig:AbsXsections} also shows the LAr emission spectrum taken from\,\cite{Neumeier:2012cz} plotted in arbitrary units (grey). The spectrum follows closely a Gaussian centred around 127\,nm with a FWHM of 7.5\,nm.  

From the data in Figure\,\ref{fig:AbsXsections} we determine ``effective'' absoprtion cross sections for LAr VUV scintillation \sigeff\ from the overlap integral of the spectral cross sections $\sigma (\lambda)$ with the (normalised) LAr emission spectrum $\Phi (\lambda)$:
\begin{equation}
\sigeff\,=\,\int \sigma(\lambda+7\,{\rm nm})~\Phi(\lambda)~d\lambda.
\label{equ:conv}
\end{equation}
Based on the observed blue shift of the Xe spectrum and the the further evidence from Hg embedded in dense argon (the 184.5\,nm line was found to be shifted to 178.5\,nm \,\cite{doi:10.1080/00268975900100401,doi:10.1080/00268976000100351,:/content/aip/journal/jcp/61/6/10.1063/1.1682342}) we apply a blue shift of 7\,nm to all absorption spectra for the calculation of the integral, corresponding to a small energy shift due to the band structure of LAr. 
\begin{table}[htb]
\centering
\begin{tabular}{c | c | c c | c}
%\hline
Molecule  & \sigeff\ [Mbarn ] &  blue shifted  & red shifted & \creq\ [ppb]\\
\hline
NO$_{2}$    &14   &12                        &16          &67    \\
\CHfour\     &9.8     &1.8                &18          &97    \\
H$_{2}$O         &4.4      &1.1            &7.8         &220   \\
O$_{2}$        &5.5       &13              &6.7         &170   \\
SO$_{2}$       &9.6        &3.8           &52          &99    \\
C$_{2}$H$_{2}$  &42       &7.8          &32          &23    \\
NH$_3$       &10            &7.3         &12          &94     \\
CO$_{2}$      &0.62          &0.53       &0.34        &1.5k   \\
NO           &2.5           &2.4          &2.2         &380    \\
\NN          &0.007           &  &                 & 135k \\
Xe             &35               &          &     &27     \\
\hline
\hline
\end{tabular}
\caption{Effective molecular absorption cross sections (second column) calculated by Equation\,\ref{equ:conv} using 7\,nm blue shifted spectra of Figure\,\ref{fig:AbsXsections}. The third and fourth columns show the values for \sigeff\ if the spectra are further shifted by 7\,nm to the blue, and to the red. The last column displays the required concentration of the molecules \creq\ to produce an attenuation length of 50\,cm. The values for \NN\ and Xe were taken from\,\cite{1748-0221-8-07-P07011} and\,\cite{AndreasUlrich}} 
\label{tab:cs}
\end{table}
We also shift the spectra for further 7\,nm to the blue and to the red to determine the sensitivity of \sigeff\ due to local spectral fluctuations. Table\,\ref{tab:cs} shows the results together with the required concentration \creq\ of the impurities to produce a 50\,cm attenuation length. The values for \NN\ and Xe could not be calculated from spectra (due to missing data), but were taken from references\,\cite{1748-0221-8-07-P07011} and \cite{PhysRevA.46.149} respectively. Molecules with cross sections in the range of 10\,Mbarn are potential candidates for the short attenuation length found in this work. In the following chapter we discuss results derived from the lifetime of the slow scintillation component, as well as measurements of the argon purity by means of trace analysers.  

%%%%%%%%%%%%%%%%%%%%%%%%%%%%%%%%%%%%%%%%%%%%%%%%%%%%%%%%%%%%%%%%%%%%%%%%%%
\section{In-situ measurements on the bulk argon} 
\label{sec:bulkargon}

The lifetime \tslow\ of the slow scintillation component can be used as an indicator for the presence of impurities in argon. In liquid argon the effect of impurity quenching  was studied in detail for the elements O$_{2}$, \NN\,\cite{1748-0221-5-05-P05003,1748-0221-5-06-P06003,Acciarri2009169} and \CHfour\,\cite{1748-0221-8-12-P12015}. Measurements done in gaseous argon\,\cite{1748-0221-3-02-P02001,1748-0221-6-08-P08003} show evidence that quenching effects of O$_{2}$ and H$_{2}$O are roughly equivalent.  

\begin{figure}[htb]
\centering
\includegraphics[height=0.5\textwidth]{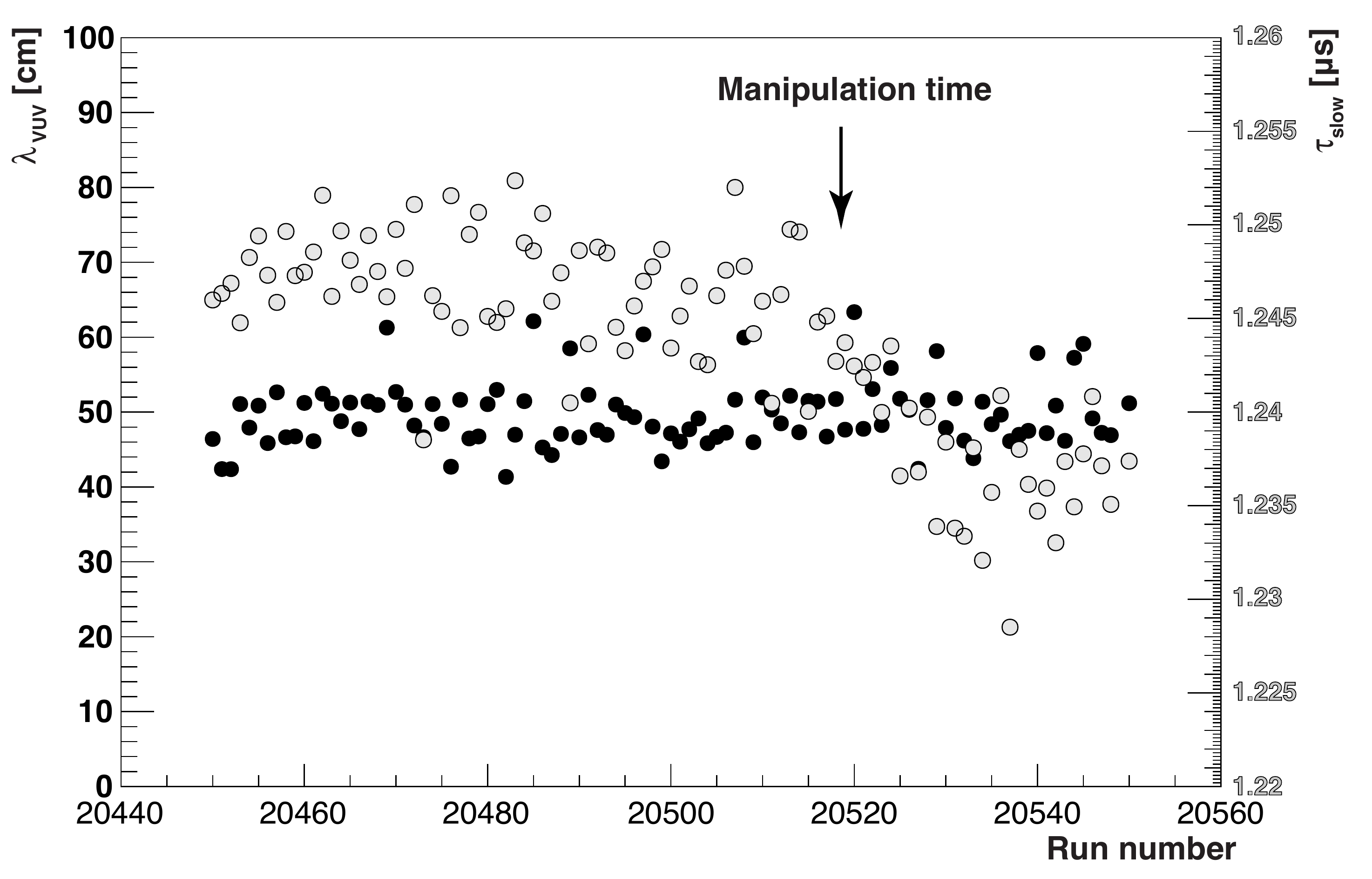}
\caption{Run by run estimates of \VUVabsL\ over a range of 100 runs (black dots, left scale). The open circles (right scale) show the corresponding values for \tslow\ which are affected by a experimental manipulation injecting some traces of \HHO\ (see text).}
\label{fig:tau3_scan}
\end{figure}

In \ardm\ the slow scintillation component was continuously monitored over the period of data taking and was stable to a level of a few percent. This was also true during the filling of the LAr target, a strong hint for introducing the impurities together with the argon gas. Figure\,\ref{fig:tau3_scan} shows the fit results for \VUVabsL\ (black dots) over a period of about 100 runs around the time when a manipulation on the gaseous recirculation system was undertaken (arrow).     
A small decrease of about 1\%  is found in \tslow\ (open circles)  before the value restores again to about 1.25\mus .  The decrease is presumably due to the injection of traces of water molecules stemming from outgassing warm detector components. On the scale of the statistical fluctuations no significant effect on the value of \VUVabsL\ can be observed. The measured value of $\tslow\approx1.25\mus$ however is about 25\% smaller than the literature value of 1.6\mus\,\cite{Hitachi:1983zz} which is commonly regarded as a good estimate for the natural lifetime of the triplet excimer state. This observation gives further evidence for the presence of impurities in the argon and quenching the light yield by about 16\% for electron like events. From \tslow\,=\,1.25\mus\ found in this work and \cite{Acciarri2009169,1748-0221-8-12-P12015} we estimate upper limits of 100\,ppb for the  concentrations of \OO , \HHO\ and \CHfour\ impurities, and about 1\,ppm for \NN\ in the \ardm\ LAr target of Run\,I. 

In addition to the analyses of pulse shapes the levels of \OO\ and \NN\ in the LAr target of ArDM were directly measured by means of two dedicated trace analysers. For oxygen the model AMI 2001RS\footnote{\url{http://www.amio2.com}} was employed, based on an electrochemical oxygen sensor ranging from 10\,ppm to 25\% in the concentration. Selecting the most sensitive range, a sensitivity of about 0.5\% of the full scale can be reached, i.e.~50\,ppb. 

The nitrogen trace in argon analyser, GOW-MAC 1200 Series\footnote{\url{http://www.gow-mac.com}}, is based on a spectral measurement of glow discharges by means of a PMT. It has two measurement ranges: 0--10\,ppm and 0--100\,ppm, with accuracies of $\pm$0.25\,ppm and $\pm$2.5\,ppm, respectively. 

Prior to measurements the two analysers were calibrated using premixed gases, 10 and 100\,ppm oxygen and 50\,ppm nitrogen in argon. The calibrations were verified flowing Ar-5.0 grade (99.999\% pure) gaseous argon, where ppm level of traces were measured for both, oxygen and nitrogen, as expected. For each of oxygen and nitrogen, the measurements were performed taking argon gas from two different points in the ArDM system: (1) at the outlet of the LAr recirculation pump, where the sampled gas was considered to be boil-off of the liquid and (2) at the gas recirculation line, where the sampled gas was considered to be the gas stratifying on top of the LAr target in the vapour phase.  After connecting the device to the system, gas was flown for several hours through the device, which is sufficient to reach the maximal sensitivity of the instruments. 

All measured results were consistent with zero within the measurement sensitivities. Considering the intrinsic sensitivities of the devices as mentioned above, and the fluctuation of the measured values, we estimate the upper limits of 0.1 and 0.5\,ppm, for the oxygen and nitrogen concentrations, respectively. The same result was found for both, the liquid and the gaseous phase of the \ardm\ target.  

%%%%%%%%%%%%%%%%%%%%%%%%%%%%%%%%%%%%%%%%%%%%%%%%%%%%%%%%%%%%%%%%%%%%%%%%%%
\section{Mass spectrometric measurements}
\label{sec:qms}  %\subsection{Trace measurements with a ICP mass spectrometer above 100 amu}

In addition to the in-situ measurements a gas sample was taken for mass spectroscopical analyses, mainly to address impurities of higher masses ($>$80\,amu) like Kr and Xe, which as mentioned above are known for their large cross sections to the LAr scintillation light. This measurement was done in collaboration with the Ionizing Radiation and Dosimetry Group of CIEMAT with a high resolution mass spectrometer evidencing differences in the mass-to-charge ratio ($m/z$) of ionised atoms or molecules below ppb levels. 
 
The inductively coupled plasma mass spectroscopy (ICP-MS) is a popular technique used to determine the concentrations in the trace and ultra-trace range by counting the number of atoms of the element which are detected. The sample is ionised by means of a plasma source, and then the ion beam is transferred to a quadrupole $m/z$ filter. A drawback of the plasma ion source is the decomposition of many lighter molecules due to the high plasma temperature. For this reason only spectra at the larger masses could be scanned. The isotopic abundances are recorded from count rates at different $m/z$ values of the quadrupole filter. The total concentrations of elements are finally found by summing over corresponding isotopes. Concentrations down to the ppq range (pg/L) can be obtained (mainly limited by sample preparation before its examination), giving both the elemental and isotopic information with minimal measurement time ($\approx$ minutes).  

The argon analysis has been carried out introducing the gas directly in a high resolution ICP-MS Element XR (Thermo Scientific) with double focusing reverse Nier-Johnson geometry and SEM/Faraday detectors. Several runs have been undertaken. In particular the \ardm\ gas, nominally of the ALPHAGAZ-2 type (N60, 99.9999\% purity), was screened and also compared to ALPHAGAZ-1 (N50, 99.999\% purity) supplied by AirLiquide Spain to CIEMAT-Madrid\,\footnote{ALPHAGAZ-1/2 Argon, Air Liquide, \url{https://www.airliquide.com/}}. The detector has been calibrated with standard samples of known concentrations and the gas has been flushed through the spectrometer in order to minimise backgrounds. 

\begin{figure}[htb]
\centering
\includegraphics[height=0.5\textwidth]{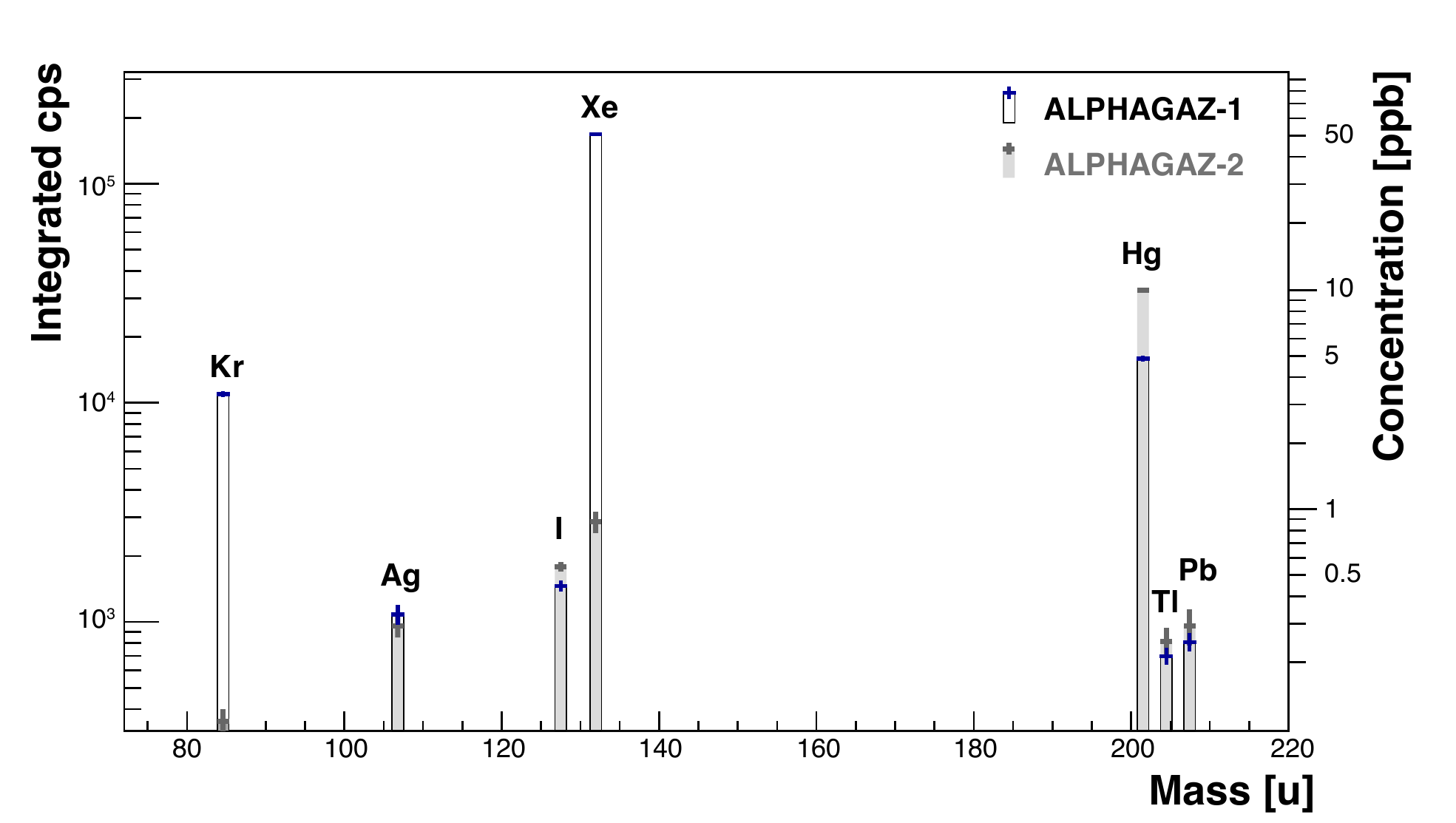}
\caption{Elements detected in the ICP mass spectral analysis for ALPHAGAZ-1 and -2. The signal strengths were calculated from the integral over all isotopes of the elements.}
\label{fig:xehg}
\end{figure}

Data with different gas fluxes have been taken by changing the input pressure of the gas. The results evidenced traces of Xe and Kr in the argon with an estimated concentration of less than 1\,ppb for the \ardm\ gas and at least one order of magnitude more for ALPHAGAZ 1. The only other trace evidenced, which is by far the most dominant (Fig.\,\ref{fig:xehg}), has been identified as Hg at the level of 10\,$\pm$\,5\,ppb. The Hg identification was surprising, but is clearly established by the relative natural abundance of different isotopes. The Hg signal has been calibrated with standard samples of known concentrations (Standard TraceCERT\textsuperscript{\textregistered} from Sigma-Aldrich, 1000\,mg/L Hg with 12\% HNO3), and the error on the measured contamination is mainly given by the uncertainty determining the argon flux through the instrument.  
It has been found that the Hg concentration is three times larger in the ArDM gas compared to ALPHAGAZ-1. A possible explanation was given by the admixture of H$_{2}$ gas in the argon production process in order to reduce the O$_{2}$ content of natural argon. The water produced in this process is later removed by a drier system. The manufacturer confirmed that the H$_{2}$ used during the process is obtained by electrolysis with Hg electrodes, so an unknown level of contamination of Hg vapour can be assumed for the H$_{2}$ gas. The higher Hg contamination found for the \ardm\ gas (ALPHAGAZ-2) in respect to the ALPHAGAZ-1, could origin in the heavier treatment of the argon with H$_{2}$ gas in order to reduce the O$_{2}$ contamination further. 

The oscillator strength of Hg for absorption in the VUV range around  127\,nm is not well explored. Similar to \NN , Hg seems to show very little optical activity in the range between the ionisation edge (~120\,nm, probably blue shifted in LAr) and the strong UV transition around 184.5\,nm (shifted to 178.5\,nm in solid argon\,\cite{doi:10.1080/00268975900100401,doi:10.1080/00268976000100351,:/content/aip/journal/jcp/61/6/10.1063/1.1682342}). An unrealistic large cross section of 100\,Mbarn would be required for 10\,ppb Hg atoms in LAr to cause an attenuation length of 0.5\,m. From the measured abundances of Kr ($<$\,0.2\,ppb) and Xe ($<$\,1\,ppb) we can also exclude heavier rare gases. For Xe a cross section of about 1GBarn would be required, about a factor 30 more than estimated by \cite{AndreasUlrich} (35\,Mbarn).

\paragraph{QMS measurements at the low mass range}

To cross check impurity levels quoted by the manufacturer we investigated a gas sample from the \ardm\ target in the laboratory by means of a quadrupole mass spectrometer type QMS 200\,M from the company PFEIFFER. The mass spectrum was also compared to argon gas of the same nominal quality (N60) from a local supplier. The system was baked (up to 150$^{\circ}$C) and pumped (below 10$^{-9}$\,mbar) for several weeks. Within the sensitivity of the instrument ($\approx$10\,ppm) no abnormal activity in the mass spectra could be observed.

%%%%%%%%%%%%%%%%%%%%%%%%%%%%%%%%%%%%%%%%%%%%%%%%%%%%%%%%%%%%%%%%%%%%%%%%%%
\section{Summary and conclusions}

The study of the VUV light yield in the tonne scale LAr detector \ardm\ resulted in a lower than expected value of 0.5\,m for the attenuation length of the liquid argon bulk to its own scintillation light. The result was found by means of a Bayesian variation technique and yielded systematic uncertainties on the order of 20\%. We interpret this result with the presence of optically-active trace impurities in the LAr which are not filtered by the installed purification systems primarily designed to target \OO\ and \HHO\ molecules. Based on the several elements studied in literature, we find evidence that optical parameters measured at room temperature in the gaseous phase, can be transferred to a liquid argon environment. However blue shifts (about 7\,nm) of the spectra and broadenings of the lines are observed. This allowed us to conduct a combined analysis of our result with respect to the involved photoabsorption cross sections, the lifetime of the slow scintillation component, as well as mass spectra taken on argon gas samples. A variety of impurity candidates could be excluded, respectively bound to upper limits in their abundance. Our data can be explained by impurities with VUV absorption cross sections in the range of 10-100\,Mbarn and concentrations of 10-100\,ppb. Due to their small cross sections, an origin from CO$_2$ and \NN\ can primarily be excluded. Their concentrations would have to be increased by several orders of magnitude to significantly influence the VUV attenuation length. Further on the measured lifetime of the slow scintillation component 1.25\mus\ constrains the concentrations of \OO\ and \HHO\ to values below 100\,ppb (as well as \NN\ to below 1\,ppm). By high resolution mass spectroscopy several heavier elements which are well known to exhibit large absorption cross sections for LAr scintillation light could also be excluded. The concentrations measured for Xe and Kr of $\leq$\,1\,ppb would require cross sections in the GBarn range to affect the VUV attenuation length in the \ardm\ setup. \CHfour\ is known to be present in argon gasses due to its abundance in the atmosphere and a similar boiling point. A quantity of 100\,ppb could explain the observed value for the attenuation length, since its VUV cross section is of the order of 10\,Mbarn. The measured slow scintillation lifetime is just about compatible with such a hypothesis\,\cite{1748-0221-8-12-P12015}.

Surprisingly also a small quantity of Hg (10\,$\pm$\,5\,ppb) was detected. Presumably Hg was introduced as a residuum during the H$_2$ cleaning cycle (to remove \OO ) in the manufacturing process of the gas. However to affect significantly the attenuation length the cross section and concentration of Hg are too low.

Elements which could not be excluded by this analysis are believed to be removable with an upgraded purification system. This is an important fact since in case of a major presence of heavier rare gases, like e.g.~Kr and Xe, the installation of a distillation system for the LAr might have been required. A cold charcoal trap is being added to the purification circuit of ArDM. Cold charcoal filters are known to reduce impurity levels of elements with condensation points above the filter temperature by several orders of magnitude. Further tests with the upgraded system will help confirm the origin of the impurities. 

The results of the work presented here has pointed out that other trace elements than the usually targeted water and oxygen molecules might affect the overall performance of a liquid argon TPC. This observation will likely have some implications on the design and optimisation of light detection systems and/or on the liquid argon purification systems of future large LAr detectors, where scintillation light attenuation length in excess of meters will be desirable.

\section{Acknowledgements}

We acknowledge the support of the Swiss National Science Foundation (SNF) and the ETH Zurich, as well as the Spanish Ministry of Economy and Competitiveness (MINECO) through the grants FPA2012-30811 and FPA2015-70657P, as well as from the "Unidad de Excelencia María de Maeztu: CIEMAT - FÍSICA DE PARTÍCULAS” through the grant MDM-2015-0509.

We thank the directorate and the personnel of the Spanish underground laboratory {\it Laboratorio Subterr\'aneo de Canfranc} (LSC) for the support of the \ardm\ experiment. We also thank CERN for continued support of \ardm\ as the CERN Recognized RE18 project, where part of the R\&D and data analysis were conducted. Further on we thank the Radiation and Dosimetry Group of CIEMAT for conducting the ICP mass spectra of the argon sample. We thank PD Dr.~A.~Ulrich (TUM) for discussions and the estimation of the molecular cross section of Xe atoms in LAr. We also thank L.~Gerchow (ETHZ) for his help with the QMS measurements at the low mass range.

%% `Bib style
\bibliographystyle{JHEP}
\bibliography{ardmbib}{}

\providecommand{\href}[2]{#2}\begingroup\raggedright\begin{thebibliography}{10}

\bibitem{10.2307/77705}
B.~Raz and J.~Jortner, \emph{Experimental evidence for trapped exciton states
  in liquid rare gases}, {\emph{Proceedings of the Royal Society of London.
  Series A, Mathematical and Physical Sciences} {\bf 317} (1970) 113--131}.

\bibitem{Suzuki79}
M.~Suzuki and S.~Kubota, \emph{Mechanism of propotional scintillation in argon,
  krypton and xenon}, {\emph{Nucl. Inst. Meth.} {\bf 164} (1979) 197--199}.

\bibitem{:/content/aip/journal/jcp/91/3/10.1063/1.457108}
E.~Morikawa, R.~Reininger, P.~Guertler, V.~Saile and P.~Laporte, \emph{Argon,
  krypton, and xenon excimer luminescence: From the dilute gas to the condensed
  phase}, \href{http://dx.doi.org/http://dx.doi.org/10.1063/1.457108}{\emph{The
  Journal of Chemical Physics} {\bf 91} (1989) 1469--1477}.

\bibitem{0295-5075-91-6-62002}
T.~Heindl, T.~Dandl, M.~Hofmann, R.~Kr\"{u}cken, L.~Oberauer, W.~Potzel et~al.,
  \emph{The scintillation of liquid argon}, {\emph{EPL (Europhysics Letters)}
  {\bf 91} (2010) 62002}.

\bibitem{:/content/aip/journal/jcp/57/8/10.1063/1.1678779}
A.~Gedanken, J.~Jortner, B.~Raz and A.~Sz\"{o}ke, \emph{Electronic energy
  transfer phenomena in rare gases},
  \href{http://dx.doi.org/http://dx.doi.org/10.1063/1.1678779}{\emph{The
  Journal of Chemical Physics} {\bf 57} (1972) 3456--3469}.

\bibitem{1748-0221-5-05-P05003}
R.~Acciarri et~al., \emph{Oxygen contamination in liquid argon: combined
  effects on ionization electron charge and scintillation light},
  {\emph{Journal of Instrumentation} {\bf 5} (2010) P05003}.

\bibitem{1748-0221-3-02-P02001}
C.~Amsler et~al., \emph{Luminescence quenching of the triplet excimer state by
  air traces in gaseous argon}, {\emph{Journal of Instrumentation} {\bf 3}
  (2008) P02001}.

\bibitem{1748-0221-5-06-P06003}
R.~Acciarri, M.~Antonello, B.~Baibussinov, M.~Baldo-Ceolin, P.~Benetti,
  F.~Calaprice et~al., \emph{Effects of nitrogen contamination in liquid
  argon}, {\emph{Journal of Instrumentation} {\bf 5} (2010) P06003}.

\bibitem{Acciarri2009169}
R.~Acciarri, M.~Antonello, B.~Baibussinov, M.~Baldo-Ceolin, P.~Benetti,
  F.~Calaprice et~al., \emph{Effects of nitrogen and oxygen contaminations in
  liquid argon},
  \href{http://dx.doi.org/http://dx.doi.org/10.1016/j.nima.2009.03.142}{\emph{Nuclear
  Instruments and Methods in Physics Research Section A: Accelerators,
  Spectrometers, Detectors and Associated Equipment} {\bf 607} (2009) 169 --
  172}.

\bibitem{1748-0221-8-12-P12015}
B.~J.~P. Jones, T.~Alexander, H.~O. Back, G.~Collin, J.~M. Conrad, A.~Greene
  et~al., \emph{The effects of dissolved methane upon liquid argon
  scintillation light}, {\emph{Journal of Instrumentation} {\bf 8} (2013)
  P12015}.

\bibitem{Neumeier201570}
A.~Neumeier, T.~Dandl, A.~Himpsl, M.~Hofmann, L.~Oberauer, W.~Potzel et~al.,
  \emph{Attenuation measurements of vacuum ultraviolet light in liquid argon
  revisited},
  \href{http://dx.doi.org/http://dx.doi.org/10.1016/j.nima.2015.07.051}{\emph{Nuclear
  Instruments and Methods in Physics Research Section A: Accelerators,
  Spectrometers, Detectors and Associated Equipment} {\bf 800} (2015) 70 --
  81}.

\bibitem{1748-0221-8-07-P07011}
B.~J.~P. Jones, C.~S. Chiu, J.~M. Conrad, C.~M. Ignarra, T.~Katori and
  M.~Toups, \emph{A measurement of the absorption of liquid argon scintillation
  light by dissolved nitrogen at the part-per-million level}, {\emph{Journal of
  Instrumentation} {\bf 8} (2013) P07011}.

\bibitem{RAZ1970511}
B.~Raz and J.~Jortner, \emph{Wannier type impurity excited states in liquid
  rare gases},
  \href{http://dx.doi.org/http://dx.doi.org/10.1016/0009-2614(70)85029-1}{\emph{Chemical
  Physics Letters} {\bf 4} (1970) 511 -- 514}.

\bibitem{Neumeier:2012cz}
A.~Neumeier, M.~Hofmann, L.~Oberauer, W.~Potzel, S.~Schonert, T.~Dandl et~al.,
  \emph{{Attenuation of vacuum ultraviolet light in liquid argon}},
  \href{http://dx.doi.org/10.1140/epjc/s10052-012-2190-z}{\emph{Eur. Phys. J.}
  {\bf C72} (2012) 2190}.

\bibitem{Rubbia:2005ge}
A.~Rubbia, \emph{{ArDM: A Ton-scale liquid Argon experiment for direct
  detection of dark matter in the universe}},
  \href{http://dx.doi.org/10.1088/1742-6596/39/1/028}{\emph{J. Phys. Conf.
  Ser.} {\bf 39} (2006) 129--132},
  [\href{http://arxiv.org/abs/hep-ph/0510320}{{\tt hep-ph/0510320}}].

\bibitem{Calvo:2015uln}
{\scshape ArDM} collaboration, J.~Calvo et~al., \emph{{Status of ArDM-1t: First
  observations from operation with a full ton-scale liquid argon target}},
  \href{http://arxiv.org/abs/1505.02443}{{\tt 1505.02443}}.

\bibitem{thedetectorpaper}
{The ArDM Collaboration}, \emph{The \ardm\ liquid argon time projection chamber
  at the canfranc underground laboratory: a ton-scale detector for dark matter
  searches}, {\emph{In preparation} (2016) }.

\bibitem{Geant4}
``{The GEANT4 (GEometry ANd Tracking) package}.''

\bibitem{Hitachi:1983zz}
A.~Hitachi, T.~Takahashi, N.~Funayama, K.~Masuda, J.~Kikuchi and T.~Doke,
  \emph{{Effect of ionization density on the time dependence of luminescence
  from liquid argon and xenon}},
  \href{http://dx.doi.org/10.1103/PhysRevB.27.5279}{\emph{Phys. Rev.} {\bf B27}
  (1983) 5279--5285}.

\bibitem{PhysRevLett.33.1365}
J.~W. Keto, R.~E. Gleason and G.~K. Walters, \emph{Production mechanisms and
  radiative lifetimes of argon and xenon molecules emitting in the
  ultraviolet},
  \href{http://dx.doi.org/10.1103/PhysRevLett.33.1365}{\emph{Phys. Rev. Lett.}
  {\bf 33} (Dec, 1974) 1365--1368}.

\bibitem{Lippincott:2008ad}
W.~H. Lippincott, K.~J. Coakley, D.~Gastler, A.~Hime, E.~Kearns, D.~N. McKinsey
  et~al., \emph{{Scintillation time dependence and pulse shape discrimination
  in liquid argon}}, \href{http://dx.doi.org/10.1103/PhysRevC.81.039901,
  10.1103/PhysRevC.78.035801}{\emph{Phys. Rev.} {\bf C78} (2008) 035801},
  [\href{http://arxiv.org/abs/0801.1531}{{\tt 0801.1531}}].

\bibitem{1742-6596-375-1-012019}
C.~Regenfus, Y.~Allkofer, C.~Amsler, W.~Creus, A.~Ferella, J.~Rochet et~al.,
  \emph{Study of nuclear recoils in liquid argon with monoenergetic neutrons},
  {\emph{Journal of Physics: Conference Series} {\bf 375} (2012) 012019}.

\bibitem{Doke:2002}
T.~Doke et~al., \emph{{Absolute Scintillation Yields in Liquid Argon and Xenon
  for Various Particles}}, {\emph{Jpn. J. Appl. Phys.} {\bf Vol. 41} (2002)
  1538}.

\bibitem{Grace15}
E.~Grace and J.~A. Nikkel, \emph{Index of refraction, rayleigh scattering
  length, and sellmeier coefficients in solid and liquid argon and xenon},
  {\emph{http://arxiv.org/abs/1502.04213} (2015) }.

\bibitem{Bettini07}
P.~Bettini et~al., \emph{Measurement of the specific activity of 39ar in
  natural argon}, {\emph{arXiv:astro-ph/0603131v2} (2007) }.

\bibitem{Konopinski66}
E.~J. Konopinski, \emph{The theory of beta radioactivity}, ch.~1 and 7.
\newblock Oxford University Press, 1966.

\bibitem{Daniel68}
H.~Daniel, \emph{Shape of beta-ray spectra}, {\emph{Reviews of modern physics}
  {\bf 40} (968) }.

\bibitem{Keefer04}
G.~Keefer and A.~Piepke, \emph{Beta specra for 39ar, 85kr, 210bi},
  {\emph{University of Alabama} (2004) }.

\bibitem{Caldwell20092197}
A.~Caldwell, D.~Kollar and K.~Kr\"oninger", \emph{"\{BAT\}, the bayesian
  analysis toolkit},
  \href{http://dx.doi.org/http://dx.doi.org/10.1016/j.cpc.2009.06.026}{\emph{Computer
  Physics Communications} {\bf 180} (2009) 2197 -- 2209}.

\bibitem{AndreasUlrich}
A.~Ulrich. TU Munich, Germany, private communication.

\bibitem{PhysRevA.46.149}
W.~F. Chan, G.~Cooper, X.~Guo, G.~R. Burton and C.~E. Brion, \emph{Absolute
  optical oscillator strengths for the electronic excitation of atoms at high
  resolution. iii. the photoabsorption of argon, krypton, and xenon},
  \href{http://dx.doi.org/10.1103/PhysRevA.46.149}{\emph{Phys. Rev. A} {\bf 46}
  (Jul, 1992) 149--171}.

\bibitem{spectral_atlas}
``{The MPI-Mainz UV/VIS Spectral Atlas of Gaseous Molecules of Atmospheric
  Interest}.'' \url{www.uv-vis-spectral-atlas-mainz.org}.

\bibitem{doi:10.1080/00268975900100401}
M.~J. McCarty and G.~Wilse~Robinson, \emph{Environmental perturbations on
  foreign atoms and molecules in solid argon, krypton and xenon},
  \href{http://dx.doi.org/10.1080/00268975900100401}{\emph{Molecular Physics}
  {\bf 2} (1959) 415--430},
  [\href{http://arxiv.org/abs/http://dx.doi.org/10.1080/00268975900100401}{{\tt
  http://dx.doi.org/10.1080/00268975900100401}}].

\bibitem{doi:10.1080/00268976000100351}
G.~W. Robinson, \emph{Discrete sites in liquids},
  \href{http://dx.doi.org/10.1080/00268976000100351}{\emph{Molecular Physics}
  {\bf 3} (1960) 301--303},
  [\href{http://arxiv.org/abs/http://dx.doi.org/10.1080/00268976000100351}{{\tt
  http://dx.doi.org/10.1080/00268976000100351}}].

\bibitem{:/content/aip/journal/jcp/61/6/10.1063/1.1682342}
S.~A. Malo, \emph{{VUV absorption spectra of mercury atoms trapped in solid
  inert gases}},
  \href{http://dx.doi.org/http://dx.doi.org/10.1063/1.1682342}{\emph{The
  Journal of Chemical Physics} {\bf 61} (1974) 2408--2411}.

\bibitem{1748-0221-6-08-P08003}
K.~Mavrokoridis, R.~G. Calland, J.~Coleman, P.~K. Lightfoot, N.~McCauley, K.~J.
  McCormick et~al., \emph{Argon purification studies and a novel liquid argon
  re-circulation system},
  \href{http://dx.doi.org/10.1088/1748-0221/6/08/P08003}{\emph{Journal of
  Instrumentation} {\bf 6} (2011) P08003}.

\end{thebibliography}\endgroup

\end{document}